\documentclass[useAMS,usenatbib]{mn2e}

\usepackage{graphicx}

\title[The influence of gravitational lensing on the spectra of
lensed QSOs]{The influence of gravitational lensing on the spectra of    
lensed QSOs}

\author[L.\v C. Popovi\'c \& G. Chartas]{L.\v
C. Popovi\'c$^{1,2}$\thanks{E-mail:lpopovic@aob.bg.ac.yu; lpopovic@aip.de} 
and G. Chartas$^{3}$\\
$^1$Astronomical Observatory, Volgina 7, 11160 Belgrade
74, Serbia \\
$^2$Astrophysikalisches Institut Potsdam, An der Sternwarte 16, 14482
Potsdam, Germany \\
$^3$Astronomy and Astrophysics Department, Pennsylvania State University,
University Park, PA 16802 }
\begin{document}

\date{Accepted 2004 Received 2004 ; in original
form 2004 July 8}

\pagerange{\pageref{firstpage}--\pageref{lastpage}} \pubyear{2004}

\maketitle

\label{firstpage}

\begin{abstract} We consider the influence of (milli/micro)lensing on the
spectra of lensed QSOs. We propose a method for the observational detection
of microlensing in the spectra of lensed QSOs and apply it to
the spectra of the three lensed QSOs  (PG 1115+080, QSO 1413+117 and 
QSO 0957+561) observed with Hubble Space Telescope (HST). 
We find that the flux ratio between images A1 and A2 of PG 1115+080 is
wavelength-dependent
and shows differential magnification between the emission lines 
and the continuum. We interpret this magnification
as arising from millilensing.
We also find that the temporal variations in the
continuum of image C of QSO 1413+117 may be caused by
microlensing, while the temporal variation observed in QSO 0957+561 was
probably an intrinsic one.
\end{abstract}


\maketitle   

\section{Introduction}

Gravitational macro and microlensing are well known phenomena that
have been widely discussed in the literature (e.g.,  Schneider, Ehlers
\& Falco 1992; Zakharov 1997; Narayan \& Betelmann 1999; Wambsganss 2001, Claeskens \& Surdej 2002 and references therein). The influence of
microlensing on the spectra (continuum and spectral lines) of lensed QSOs
has been investigated mainly theoretically (Nemiroff 1988; Schneider \&
Wambsganss 1990; Wambsganss \& Paczynski 1991; Popovi\'c et
al. 2001a; Abajas
2002; Lewis \& Ibata 2004). Some of the observational effects have also
been presented in several papers (e.g.,  Lewis et al. 1998; Mediavilla et
al. 1998; Wisotzki et al.  2003; Wucknitz et al. 2003; Chelouche 2003;
Metcalf et al. 2004; Richards et al. 2004).

The influence of microlensing on the line profile of lensed QSOs 
 was initially investigated by Nemiroff (1988) and Schneider \& Wambsganss
(1990).
Recently Wandel et al. (1999) and Kaspi et al. (2000) using reverberation
technique 
discovered that the broad line region (BLR) size was smaller than
that assumed in the standard AGN model. 
Following this discovery, the influence of microlensing on the line
profile of lensed QSOs in the UV/optical wavelength band,
was revised and discussed in several papers (Popovi\'c 2001a; Abajas
2002; Lewis \& Ibata 2004). 
Similar investigations have been performed in the X-ray
spectral region (Popovi\'c et al. 2001b;
Chartas et al. 2002,2004; Popovi\'c et al. 2003; Dai 2003).

Abajas et al. (2002) noted that the influence of microlensing on the
shapes of spectral lines can be very strong for high ionization 
lines arising from the inner part of the broad
emission line region.  The magnification depends on the size of
the BLR and we therefore expect that distortions of spectral lines 
will not be observable in all microlensing events. 
Using the BLR size estimation for  NGC~5548 (Peterson \& Wandel 1999) and
the relation between the radius of the BLR and optical luminosity,
$R_{BLR}\sim L_{\lambda}^{-0.7}$ at $\lambda=5100\ \AA $, given by  
Kaspi et al. (2000),
Abajas et al. (2002) identified a group of 10 gravitational lensed
systems in which 
microlensing of the BLR could be observed.

Observations and modeling of microlensing of the BLR 
are promising, because the study of the variations of the broad emission
line shapes in a microlensed QSO image could constrain  the size of the
BLR and the
continuum region (e.g., Lewis \& Ibata 2004). 
Observations of microlensing of the profiles of Fe K$\alpha$ spectral lines
in three multi-imaged QSOs were recently presented in Chartas et
al. (2002,2004) and Dai et al. (2003).  The observed variations of
the Fe K$\alpha$ lines are in good agreement
with theoretical predictions (Popovi\'c et al. 2001a; Popovi\'c et   
al. 2003).  Observations of the effect of microlensing on the profile of
{  broad} UV 
emission lines (C IV and  S IV/OIV) 
were recently reported by Richards et al. (2004).

The detection of microlensing from the distortion of the spectra of lensed QSOs
is more complicated than inferring its presence from photometric 
observations. First of all, larger telescopes are needed to obtain 
the high signal-to-noise (S/N) ratio spectra, preferably
with 2D spectrophotometry that can provide simultaneous spectra of all
images. In addition, the intrinsic variability of lensed QSOs
needs to be taken into account when comparing spectra of different images.
Gravitational microlensing can also distort the spectrum of the continuum.
Specifically, it is expected that microlensing is wavelength 
dependent (e.g., Wambsganss \& Paczynski 1991; Lewis et al. 1988; Wisotzki et
al. 2003) since the size of a continuum emission region of the accretion
disk
depends on the wavelength band. 

{ According to the standard model of AGNs, a QSO consists of a black
hole 
surrounded by a (X-ray and optical) continuum emitting region probably
with an accretion disk geometry,
a broad line region and a larger region that can be resolved in several
nearby
AGN that usually is referred to as the narrow line region
(e.g., Krolik 1999).

Variability studies of QSOs indicate  that the size of the X-ray
emission region is of order 10$^{14-16}$~cm (e.g., Chartas et
al. 2001; Oshima et al. 2001, Dai et al 2003).
An Fe K$\alpha$ fluorescence line detected in several AGN near 6.4~keV is
thought to originate from within a few 10s of gravitational radii (e.g.,
Tanaka et al. 1995; Fabian et al. 1995).
This line is thought  to be a fluorescence line of Fe due to
emission
from a cold or ionized accretion disk that is illuminated from a
source of hard X-rays originating near the central object.
Aspects of this emission can provide a probe of strong gravity near
a black hole.}
 
Gravitational lensing is in general achromatic (the deflection
angle of a light ray does not depend on its wavelength),
however, the wavelength dependent geometry of the
different emission regions may result in chromatic effects.
{ Studies aimed at determining the influence of microlensing on the
spectra
of lensed QSOs
need to account for the complex structure of the QSO central
emitting region. 
Since
the sizes of the emitting regions are wavelength dependent,
microlensing by stars in the lens galaxy will lead to a wavelength
dependent magnification.
The geometries of the line and the continuum emission regions
are in general different and there may be a variety of
geometries depending on the type of AGN  (i.e., spherical, disk-like,
cylindrical, etc.).

On the other hand, in some cases
the potential of the lens galaxy may be perturbed  by small satellite
galaxies or globular clusters (hereafter millilensing). These
perturbations will
add
additional complexity to the
magnification function.}

 To investigate the effects of microlensing, a method is needed that can
be applied 
to the observed spectra. A technique
that is often used (in cases where flux losses from slits 
are significant)
 relies on measuring the equivalent widths of spectral lines
(Lewis et al. 1998; Wisotzki et al. 2003). One limitation of this 
approach is that it does not account  for microlensing of the continuum
component 
that may significantly affect the line equivalent widths.

The aim of this paper is to discuss the influence of gravitational
(micro/milli)lensing on the spectra (continuum and spectral lines) of
lensed QSOs. 
First in \S 2 we outline a model that includes the geometry of the line
and continuum emission regions,
as well as the magnification due to the lens galaxy and the stars in the
lens galaxy. 
Based on this model we propose a method
for detecting microlensing events in the lines as well as in the continuum
components of QSOs. In \S 3 we
applied the method to a sample of lensed QSOs observed with HST,
and in \S 4 we present our conclusions.

\section{The influence of lensing on QSO spectra}

We assume that the 
emitting region of a QSO is geometrically complex, and that the 
 geometries and sizes of the continuum and the line emitting regions are
different, i.e., 
that the unlensed surface brightness, $I(\lambda;X,Y)$ (in 
the case of X-ray $I(\lambda;X,Y)\to I(E;X,Y)$, where $E$ is the emitted
energy), can be written as a function of wavelength as,

$$I(\lambda;X,Y)=I^c(\lambda;X,Y)+I^L(\lambda;X,Y),\eqno(1)$$
where  $I^c(\lambda;X,Y)$ and $I^L(\lambda;X,Y)$ are the surface brightnesses of the 
continuum and the lines, respectively and X,Y are the coordinates
of the emission region in the source plane.
 The magnification of the images depends in part on the projected 
mass distribution of the lens galaxy. We assume that the
macro-magnification
(A(X,Y)) is complex and can be represented as function of the X,Y
coordinates in
the source plane. Under these assumptions an observer will detect the
following magnified intensity of an image ($i$),

$$\Im_i(\lambda;X,Y)=\Im_i^c(\lambda;X,Y)+\Im_i^L(\lambda;X,Y)$$
where
$$\Im_i^{c,L}(\lambda;X,Y)= I^{c,L}(\lambda;X,Y)\cdot A(X,Y)\eqno(2)$$

In general one can expect an energy dependent magnification
of the images of a lensed QSO resulting in different magnifications
of the line and the continuum emission regions (as it was observed in the
case of Q2237+0305, see Mediavilla et al. 1998).

 Taking into account that the sizes of the continuum and the line emitting
regions
may be a function of wavelength, we write the fluxes as,

$$\Phi^{c,L}_i(\lambda)=\int_\Sigma I^{c,L}(\lambda;X,Y)\cdot
A(X,Y)d\Sigma, \eqno(3)$$
where $\Sigma$ is the projection of the
QSO emitting region in the source plane onto a plane 
perpendicular to the line-of-sight of the observer and $d\Sigma=dXdY$.
As one can see from Eq. (3), in general, the fluxes,
$\Phi^i_L(\lambda)$ and 
$\Phi^i_c(\lambda)$,  depend on the geometry of the source as well as on
the magnification caused by the lens galaxy.

If we assume there are no millilensing
and the sizes of the  emission regions are
significantly smaller than the 
size of the macro-caustic, then we expect the magnification of the
continuum and the line fluxes to be the same, i.e.,
$$\Phi^c_i(\lambda)=F^c(\lambda)\cdot A_i,\ \ \ 
\Phi^L_i(\lambda)=F^L(\lambda)\cdot A_i.\eqno(4)$$
where $A_i$ is the macro-magnifications of the image $i$ and
$F^{c,L}(\lambda)=\int_\Sigma I^{c,L}(\lambda,X,Y)d\Sigma$.

The inference of a { microlensing event (MLE)} from the analysis of
the spectral lines and
the continuum spectral components of lensed QSOs can be complicated by 
several effects that may include
intrinsic variability of their fluxes and their spectra and 
the fact that intrinsic  variability between different images of a lensed
QSO will be time-delayed. 

A possible solution to this problem is to obtain observations of all the images
 of a lensed QSO at different epochs, preferably with the time intervals
between observations 
coinciding with the time-delays of the system. 
 With such observations one can compare the flux of each component as
a function
 of wavelength for different epochs, as well as the flux ratios between
different
images observed at a given epoch.
We define the observed intensities of fluxes A and B 
as $\Phi_A(\lambda)$ and $\Phi_B(\lambda)$ at time $t_0$
and $\Phi_A'(\lambda)$ and $\Phi_B'(\lambda)$ at time $t_1$,
and the time-delay between the images as $\Delta t=t_1-t_0$. 

{ In principle, images of a lensed quasar
are always being microlensed by stars in the
foreground galaxy. This results in the continuous magnification as well
as demagnification of
the lensed quasar images. In the quiescent
periods, when the magnification is low, the magnification
map is smooth and one should not expect any detectable differences
between the spectral
energy distributions of images.
In the case where magnification caustics are in the vicinity of the
source
the differential magnification of the continuum is expected to be
significantly stronger and detectable (e.g., Wambsgauss \& Paczynski
1991). 
Consequently, hereafter we consider only high-magnification
($A_{MLE}$)  microlensing events.}

If a microlensing event { (with a strong amplification}) is present in
only one of the images   
(e.g., $A$ at time $t_1$), then the flux of this image will vary
with time, and the magnified flux due to microlensing can be written
as:
$$\Phi_i'(\lambda)=\int_\Sigma \Im_i(\lambda;X,Y)\cdot
A_{MLE}(X,Y)d\Sigma=$$
$${A_i}\int_\Sigma[I^c(\lambda;X,Y)+I^L(\lambda;X,Y)]
\cdot A_{MLE}(X,Y)d\Sigma\eqno(5)$$  
where $\Sigma$ is the projected surface of the macrolensed image at the
source
distances and $A_{MLE}$ is the amplification due to microlensing.
 The flux ratio ($R_{A',A}=\Phi_A'(\lambda)/\Phi_A(\lambda)$) of the same image at two 
different epochs separated by the time-delay is:

$$R_{A',A}={\int_\Sigma[I^c(\lambda;X,Y)+
I^L(\lambda;X,Y)]A_{MLE}(X,Y)d\Sigma\over{
[F^c(\lambda)+F^L(\lambda)]}},\eqno(6)$$ 
and the flux ratio between different images observed 
at the same epoch is
$$R_{A',B'}={A_1\over
A_2}{\int_\Sigma[I^c(\lambda;X,Y)+
I^L(\lambda;X,Y)]A_{MLE}(X,Y)d\Sigma\over{
[F^c(\lambda)+F^L(\lambda)]}}.\eqno(7)$$   

We define as an indicator of variability ($\sigma$): 

$$\sigma(\lambda)=R_{A',A}-R_{B',B}.\eqno(8)$$
In general $\sigma=0$ indicates no variability. 
We note the rare case where $\sigma = 0$ and variability is
present. This condition can occur when the fluxes of both images A 
and B
change by the same factor between epochs $t_{0}$ and $t_{1}$.
The probability of such an event is negligible.
If $\sigma\neq 0$, then we investigate the source of the variability.

If microlensing of only one image (e.g. A) is present and there 
is no other source of variability, then 
the average magnification caused by microlensing is,

$$\overline{A}_{MLE}(\lambda)={\int_\Sigma[I^c(\lambda;X,Y)+
I^L(\lambda;X,Y)]A_{MLE}(X,Y) d\Sigma\over{
[F^c(\lambda)+F^L(\lambda)]}}.$$

From Eqs. (5-7) we write the average magnification caused by microlensing
as

$$\overline{A}_{MLE}(\lambda)={\Phi_A'(\lambda)\over{\Phi_A(\lambda)}}={A_2\over
A_1}\cdot {\Phi_A'(\lambda)\over{\Phi_B'(\lambda)}},\eqno(9)$$
 where $A_{1}$ and $A_{2}$ are the macro-magnifications of images A and B,
respectively.

Eq. (9) is only valid for the case where microlensing is occurring in one
component. For the case where intrinsic variability
($\sigma_I$) is
present we have
$$\sigma_I=R_{A',A} - {A_2\over A_1} R_{A',B'}\neq 0,\eqno(10)$$

Also, $\sigma_I$ will be different from zero if both 
images are microlensed at the same time. This may be checked by comparing
the spectra of two images observed at two different epochs separated by
the time-delay. If the relation
$$R_{A'B}={\Phi_{A'}\over{\Phi_{B}}}={A_1\over{A_2}},$$
holds we conclude that intrinsic variation is present, otherwise if it is
not satisfied 
we conclude that both images are microlensed.

On the other hand, taking into account that the total flux is the 
sum of line and continuum fluxes
(see Eq. (1)), the average magnification of only the line is
$$\overline{A}^L_{MLE}(\lambda)={\Phi_{A'}(\lambda)-\Phi_{A'}^c(\lambda)
\over{\Phi_A(\lambda)-\Phi_{A}^c(\lambda)}}=$$
$$={A_2\over A_1}{
\Phi_{A'}(\lambda)-\Phi_{A'}^c(\lambda)
\over{\Phi_{B'}(\lambda)-\Phi_{B'}^c(\lambda)}},\eqno(11)$$
and in the case of millilensing one can expect that magnifications in the
line  and in the continuum are different ($A^L\ne A^c$) as well as that
these magnifications are wavelength dependent.

In Table 1 we point out several cases where our method can
be used. There are more complex cases not listed in Table 1
such as the combination of milli and microlensing, microlensing and 
intrinsic variability
and microlensing of both images.
Observations in several epochs separated by the time-delay are useful 
in distinguishing
between these complex cases.

\begin{table*}
\begin{minipage}{160mm}
 \caption{A list of cases where our proposed method can be used to infer 
the
presence of millilensing, microlensing and intrinsic variability.}
 \begin{tabular}{|l|l|}
\hline
The case & Description\\
\hline
I) $\sigma=0$ and $R_{AB}=R_{A'B'}=const.$ & No variability or
millilensing
present\\
& \\
IIa) $\sigma=0$, $R_{AB}=R_{A'B'}=f(\lambda)$ and/or $A^{L}\neq A^{c}$
&There is no variability, but millilensing  or differential extinction \\
IIb) $\sigma=0$, $R_{AB}=R_{A'B'}=const.$ and $A^{L}\neq A^{c}$& might be
present (see e.g., Wucknitz et al. 2003), if
$R_{AB}=f(\lambda)$ \\
& cannot be explained by extinction $\to$ millilensing is present\\
& \\
III) $\sigma\neq 0$, $\sigma_I\neq 0$, and $R_{A'B}={A_1\over{A_2}}$&
Intrinsic
variability is present\\
& \\ 
IVa) $\sigma\neq 0$,  $\sigma_I=0$, $R_{A,B}\neq
f(\lambda)$
and $R_{A',B'}=
f(\lambda)$& Microlensing of one image at time $t_1$\\
& \\
IVb) $\sigma\neq 0$,  $\sigma_I=0$, $R_{A,B}= f_1(\lambda)$, $R_{A',B'}=
f_2(\lambda)$ & Microlensing of one
image  at times $t_0$ and $t_1$\\
 and $R_{A'A}=1$ (or $R_{B'B}=1$) & (timescale of microlensing is longer 
than time delay)\\
\hline
 \end{tabular}
\end{minipage}
\end{table*}

\section{Observations: Several Examples, Results and Discussion}

Our method of analyzing the spectra of lensed quasars outlined in \S 2 
can be applied accurately when simultaneous observations of spectra of images 
of a lensed QSO at two different epochs separated by the time-delay
are available. 
In many cases the time-delay of a system is unknown. In such a situation
our method can still be applied to estimate the origin of the
magnification. We note that the application of our method is limited by the S/N of the 
component spectra, e.g., in the case of
low S/N, non zero values for the variability indicator sigma and 
chromatic effects
may only be detected in cases where there are large flux magnifications 
caused by
microlensing or other effects listed in Table 1.
Therefore high S/N spectra are needed to apply our model accurately.

Here we apply our method to several observations of lensed quasars 
performed with HST. By inspection of the HST database we
found only three lensed quasars with simultaneous
observations of all images at different epochs: QSOs Q0957+561, PG 
1115+080 
and Q1413+117. In this section
we compare the spectra between different images as well as between the same
image at different epochs using the method described above.
The  STARLINK  DIPSO software package (Howarth et al.
2003) was used to analyze the spectra.

\subsection{Q0957+561}

Images A and B of Q0957+561 have been observed with HST at several epochs. 
In the HST archive we found observations of both images made with the FOS
 G270 and obtained in 1995 (Jan 26, Oct 20, Nov 04, Nov 17, and Dec
14) that we used for
our analysis. We also found observations of only image A of Q0957+561 
with STIS performed on 1999 April 04,  and observations of only image B 
on 2000 June 2, and only image A on 2000 September 8, but these observations 
have not been used in our present analysis. A detailed description of 
all observations and  the analysis of the spectra
can be found in Michalitsianos et al. (1997),
Dolan et al. (2000) and  Hutchings (2003).

A first inspection shows that images A and B have
similar spectra (practically the same), with the exception of the spectra observed on
1995 January 26 (see Figure 1). We applied the method described above to all
observed spectra of Q0957+561 obtained in 1995,  
comparing the spectra of images A and B from different epochs. 
From observations taken within the period
October--December we did not detect any significant difference between  
the spectra that would indicate a 
microlensing event or intrinsic variability. A slight but not
statistically significant 
difference exists in the variability indicator $\sigma$ (see Figures 2a,
3a and 5a). A
significant difference between images was only detected in spectra
 observed on 1995 January 26, (Figures 2b, 3b and 5b). In Figure 2 we
present the 
 variability indicator between observations performed on October 20 and
December 14, 1995 (Fig. 2a) as an
example. The variability indicator  $\sigma$
(similar to the case of all spectra observed within the period October--December), 
does not significantly differ from zero and taking into account the noise in the spectra 
we conclude that significant variability is not
present in this time interval. On the other hand a comparison between
the spectra observed on 1995 January 26 and spectra taken at other epochs
shows significant differences. As an example the variability indicator
$\sigma$ between observations
performed on Jan 26 and Dec 14, 1995 is shown in Figure 2b.
Variability is clearly present in the continuum as well as in  
the Ly$\alpha$ line (the location of the line is indicated with a dashed line).  To
investigate the nature of this variability we
determined the flux ratio using Eq. 6.  As shown in Figure 3, the
flux ratio for the observed spectra in the period October--December is
practically one, whereas the
flux ratio of the spectra observed on
1995 January 26, is $\sim$ 1.8 times higher than the ratio observed during the period
October--December.

\begin{figure}
\includegraphics[width=8.5cm]{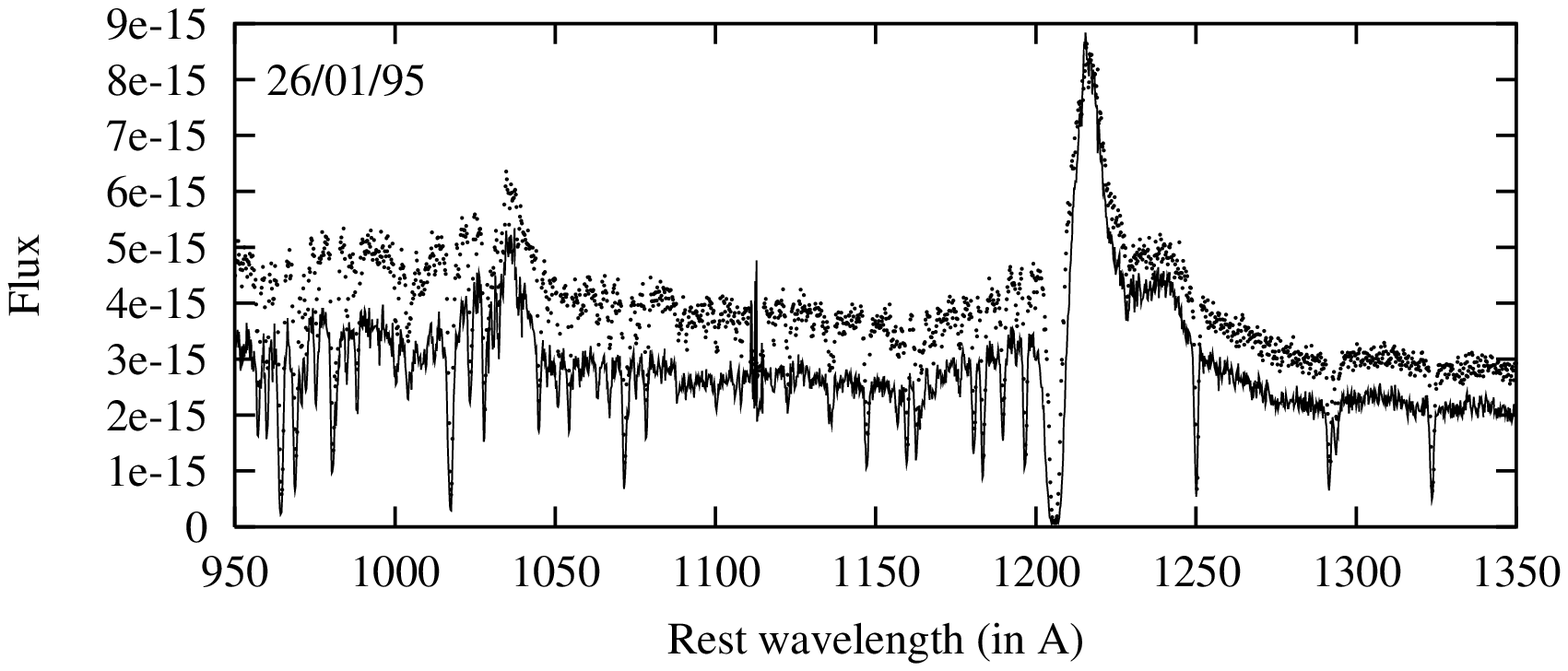}
\includegraphics[width=8.5cm]{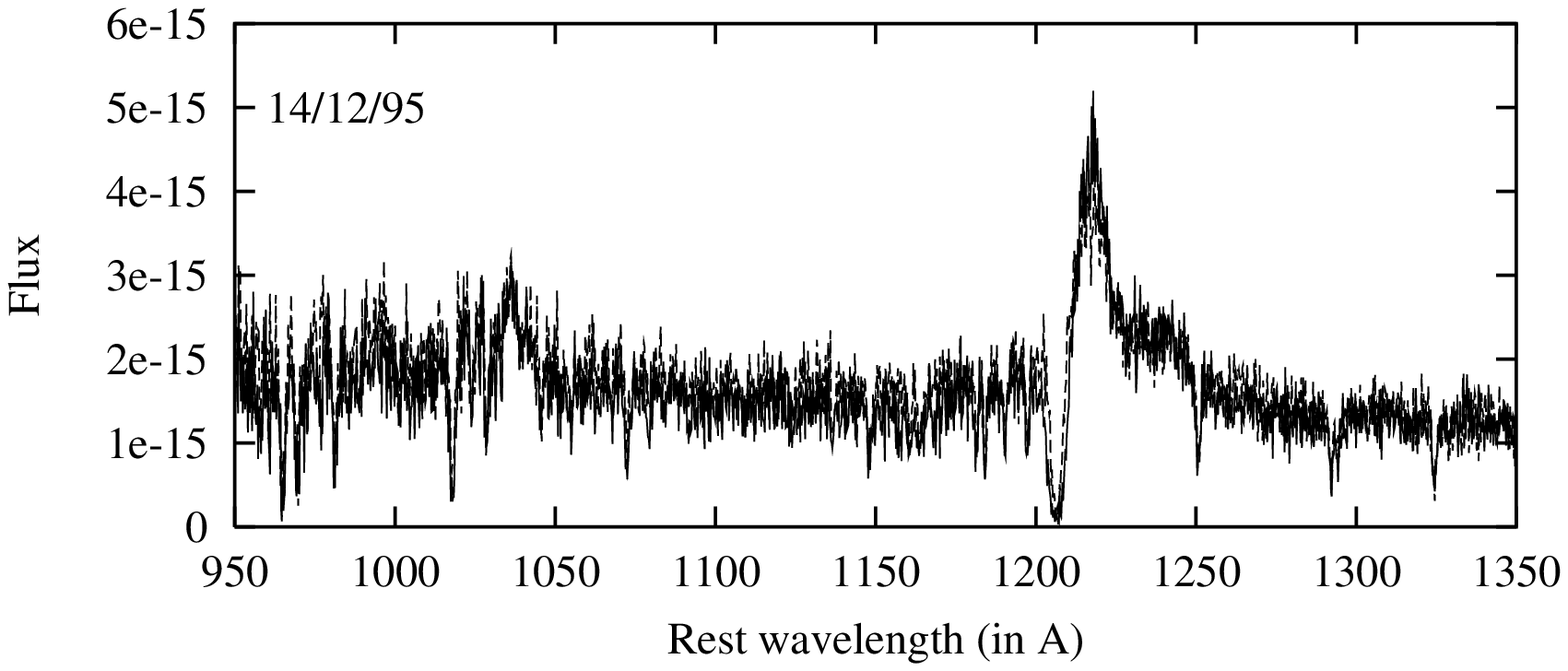}
\includegraphics[width=8.5cm]{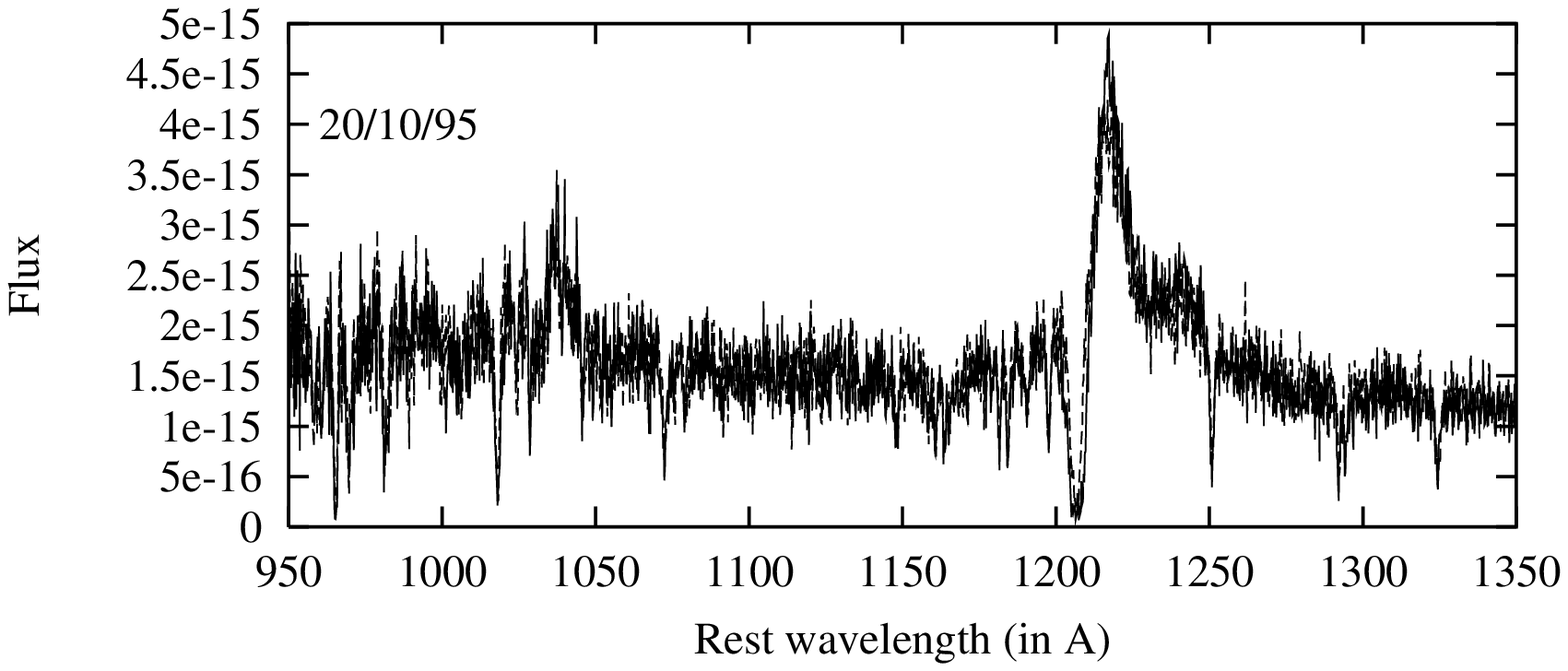}
\caption{The spectra  of Q0957+561 observed with
HST for
three different
epochs. The solid line and dots indicate the spectra of images A
and B,  respectively. The flux is given in $\rm erg\
cm^{-2}sec^{-1}A^{-1}$.}
\end{figure}

\begin{figure}
\includegraphics[width=8.5cm]{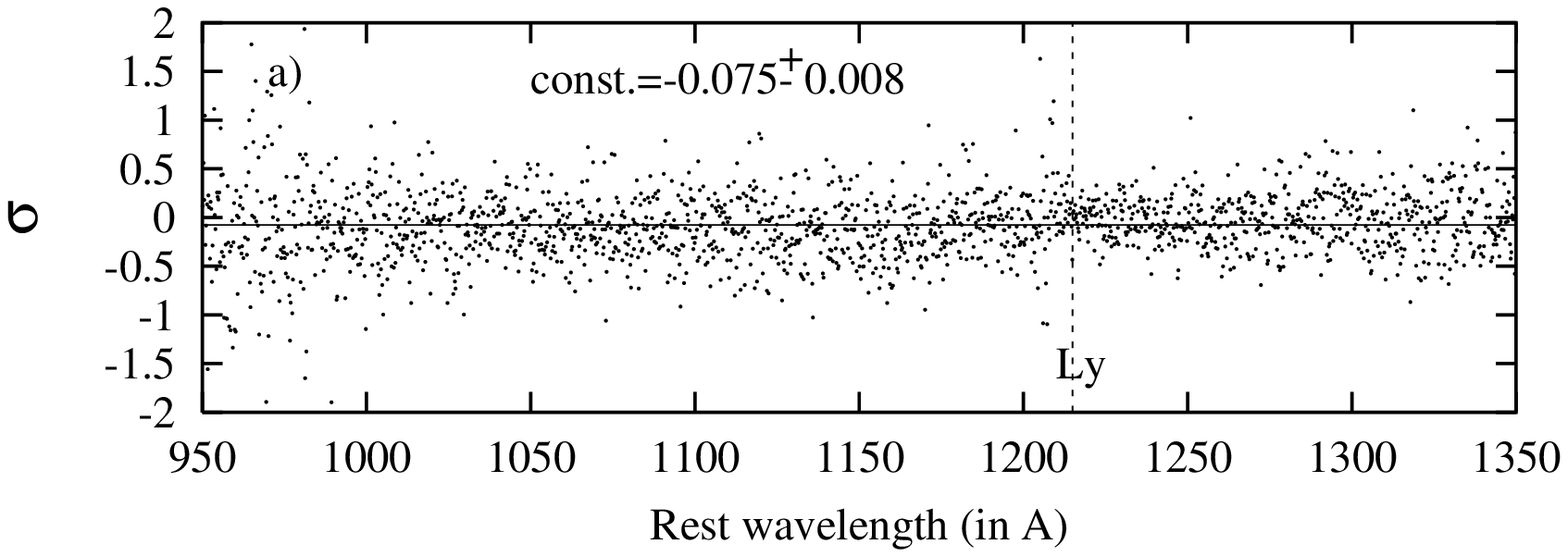}
\includegraphics[width=8.5cm]{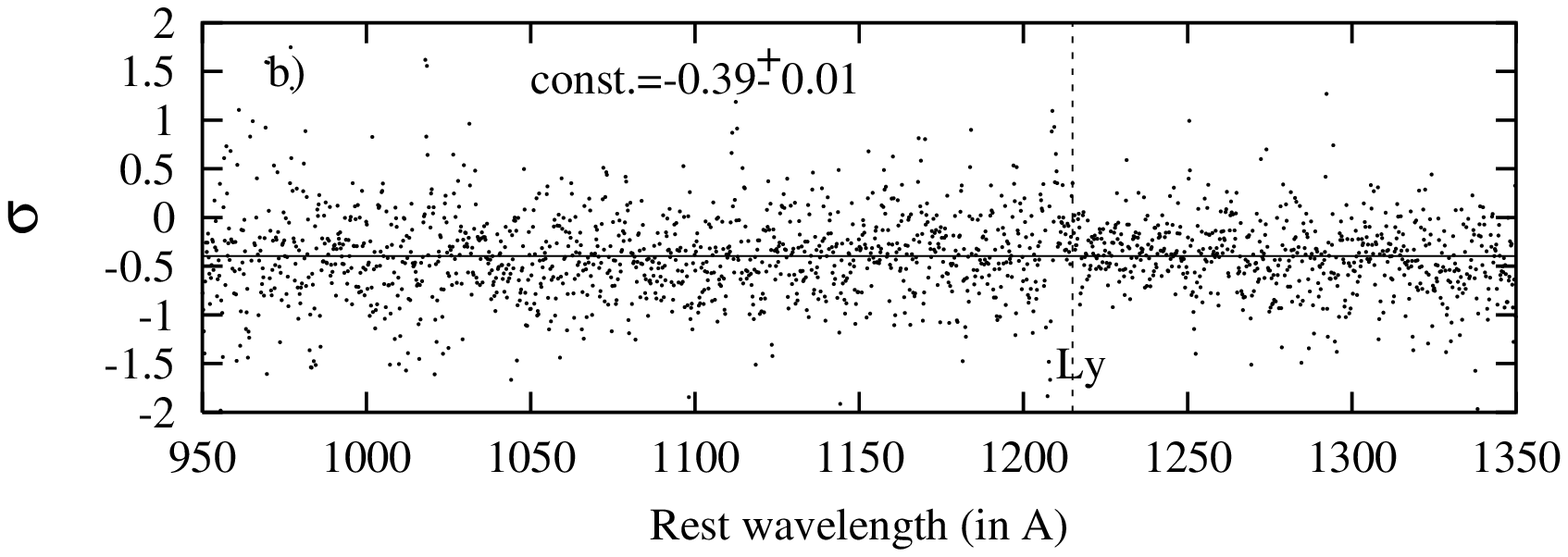}
\caption{The variability indicator $\sigma=(I_A'/I_A)-(I_B'/I_B)$ 
calculated for the spectra of Q0957+561 for the epochs: a) 1995 December 14 and
1995 October 20 (case I); b)  1995 January 26 and 1995 December 14 (case II).
The vertical dashed line indicates the location of the Ly$\alpha$ line.}
\end{figure}

\begin{figure}
\includegraphics[width=8.5cm]{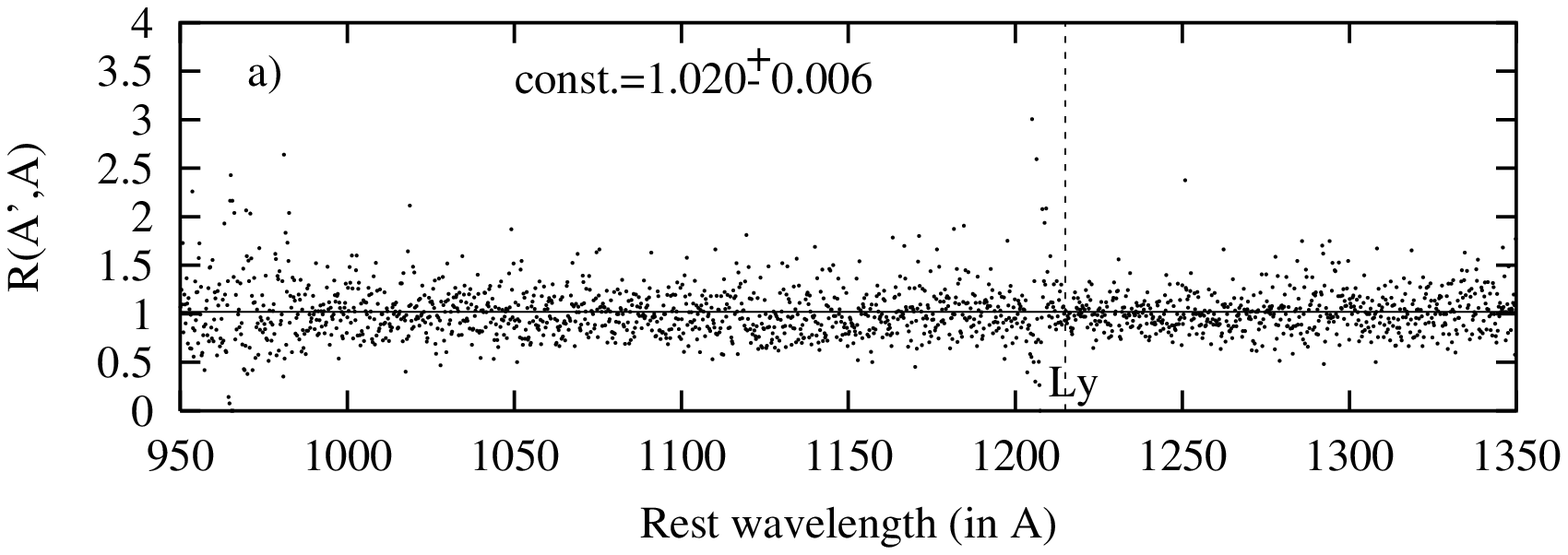}
\includegraphics[width=8.5cm]{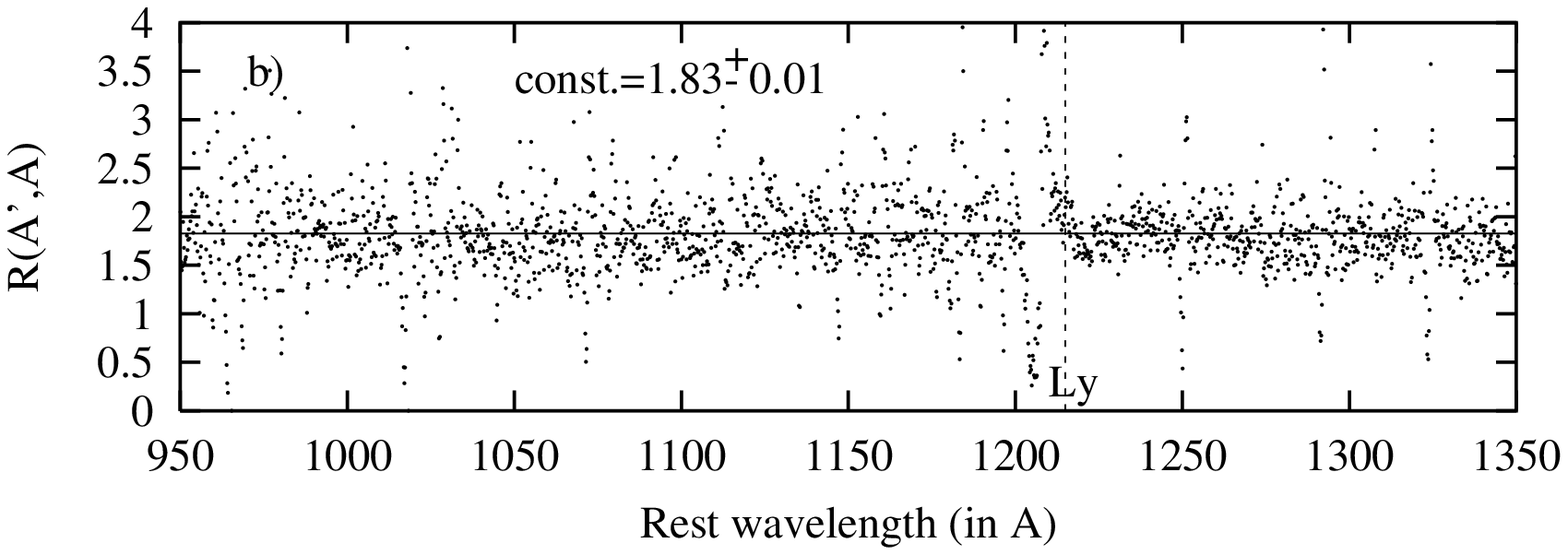}
\caption{The  flux ratio between the spectra of image A of Q0957+561
for the  epochs: a) 1995  December 14 and 1995 October 20 (case
I); b)  1995 January 26
and 1995 
December 14 (case II).}
\end{figure}

\begin{figure}
\includegraphics[width=8.5cm]{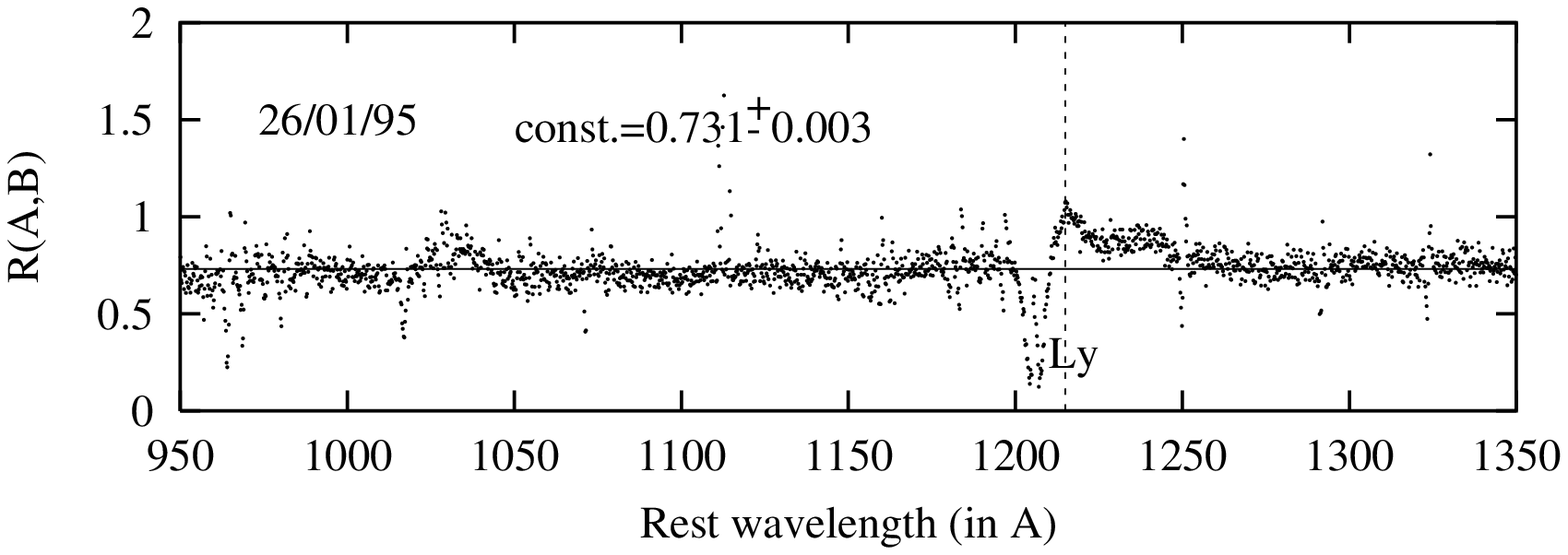}
\includegraphics[width=8.5cm]{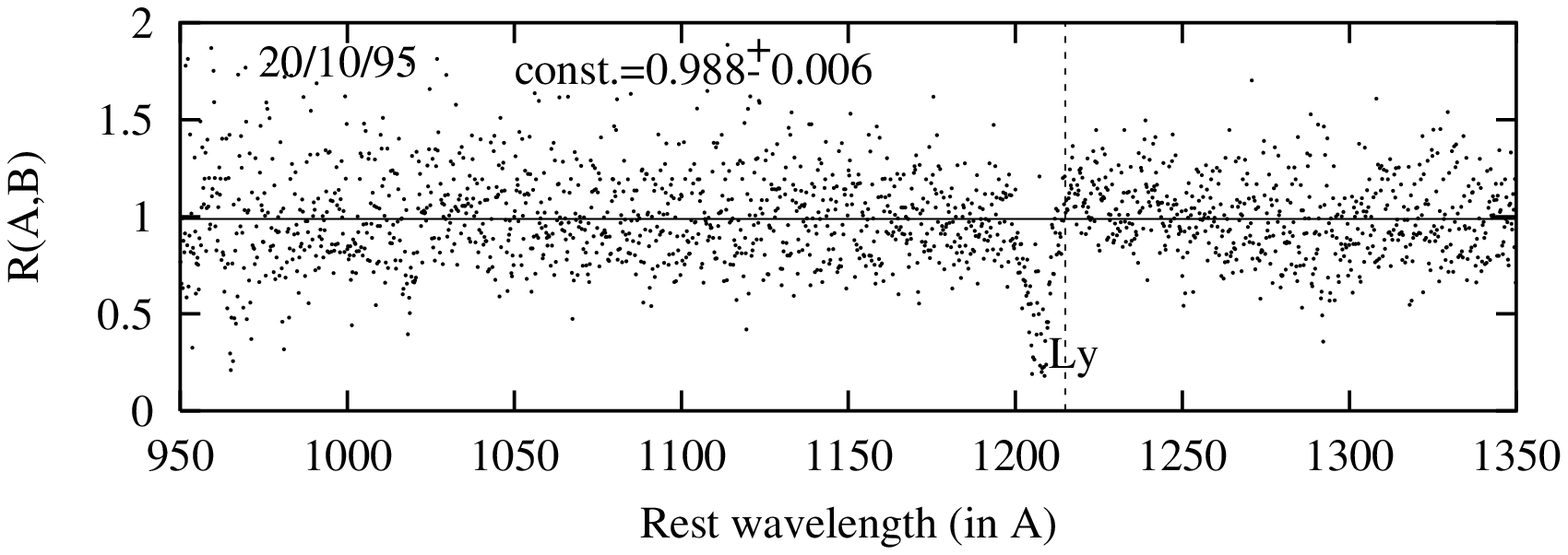}
\includegraphics[width=8.5cm]{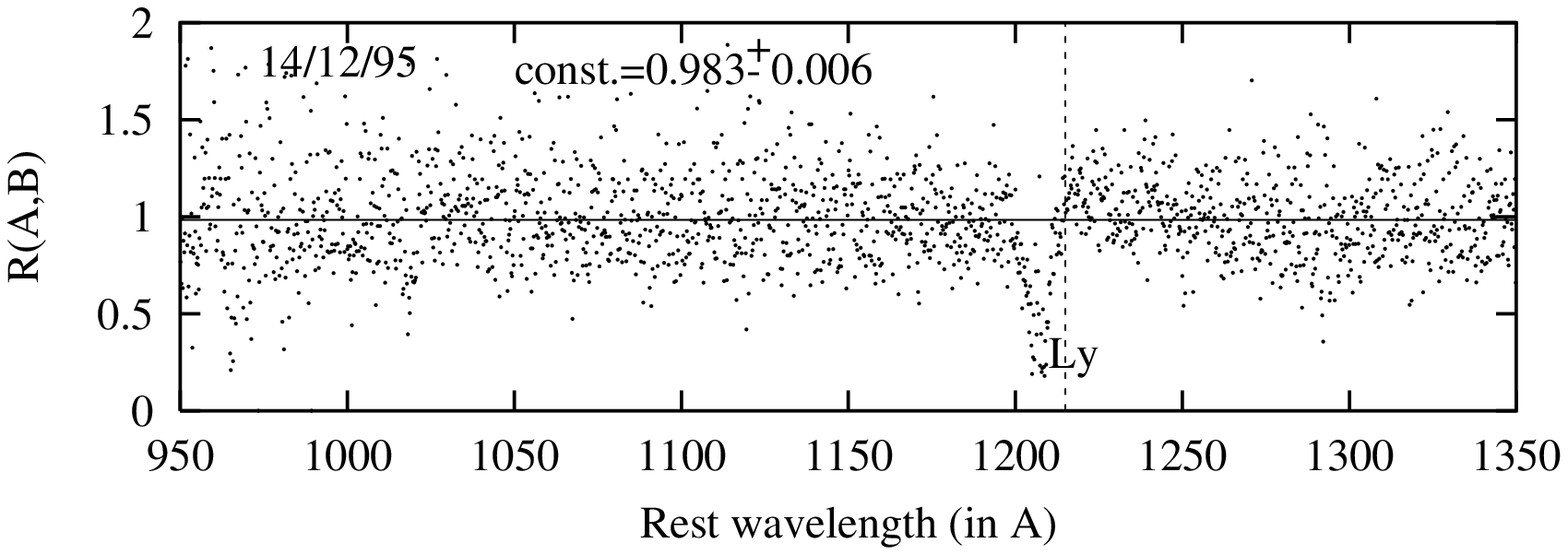}
\caption{The flux ratio between images A and B of Q0957+561 observed
for the
listed epochs.
}\end{figure}

\begin{figure}
\includegraphics[width=8.5cm]{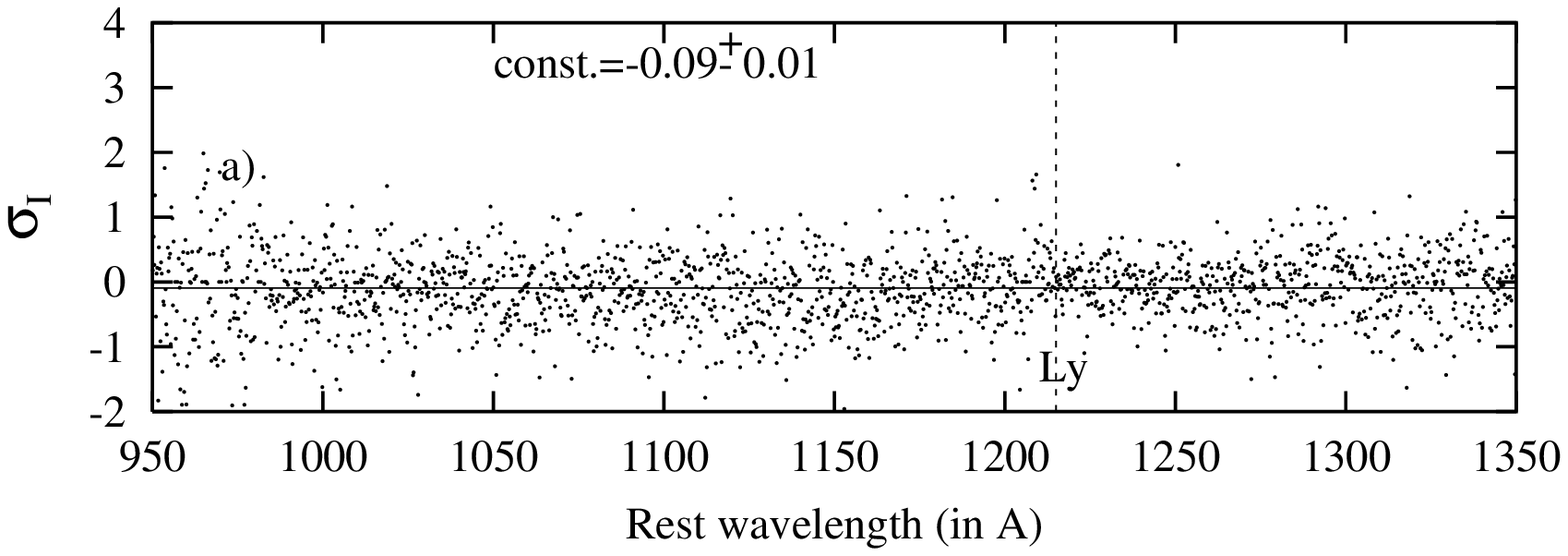}
\includegraphics[width=8.5cm]{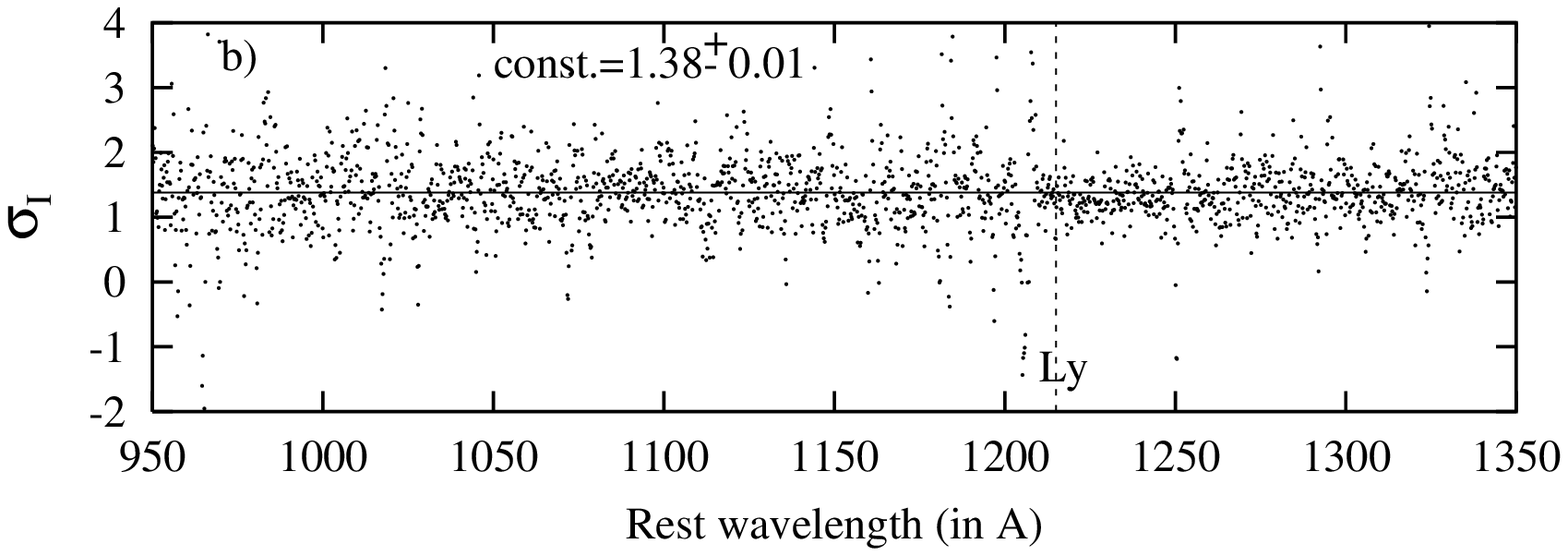}   
\caption{$\sigma_I$, an indicator of intrinsic variability of
Q0957+561
spectra
is shown for the epochs: a)   October and December
1995 and b)   January and December
1995.}\end{figure}

We do not find any significant wavelength-dependence 
of the variability indicator $\sigma$, the flux ratio $R_{A,A'}$
between the same image at different epochs 
or the flux ratio $R_{A,B}$ between different images for the same epoch.

Moreover, in Figure 4 we show that at three different epochs within the
October--December period the flux ratio $R_{A,B}$ 
is consistent with the magnification resulting from macrolensing.
Using the optical magnitudes of images A and B of Q0957+561 available
from the {\sl Castles} database\footnote{http://cfa-www.harvard.edu/castles/}
we obtained a flux ratio of $F_A\approx 1.08F_B$.
A comparison of the flux ratios of Figure 4 also shows
 that on 1995 January 26, the flux of the continuum of image B was
significantly larger 
than that of image A.

We use Eq. (10)\footnote{Here we
approximate the amplification ratio as ${A_2\over A_1}={A_{B}\over A_{A}}
\approx R_{B,A}$}
to investigate the nature of the variability in Q0957+561.
As shown in Figure 5, the
variability ($\sigma_I\neq 0$) seen on January 26, 1995  is most likely
intrinsic.

Here we briefly discuss the absorption line detected blue-wards of
Ly$\alpha$ originating from a damped Lyman alpha (DLA) system at a 
redshift of $\sim$ 1.391 (Michalitsianos 1997). 
A comparison of the flux ratios between images
at all epochs indicates that the absorption line is stronger in image A than in
image B (see Figure 4). This is in agreement with the previous analysis reported by
Michalitsianos et al. (1997). Such differences in the absorption between images indicates 
inhomogeneity in the DLA system that may arise from 
the different paths of the rays of the two images (see e.g. Chelouche 2003).

To summarize, our analysis of the spectra of Q0957+561
taken in January, October and December of 1995 indicates that:
i) the strong magnification detected in the continuum and in
the Ly$\alpha$ line is most likely caused by intrinsic variability
of the QSO,
ii)  during this variation image A has a stronger line
flux than image B and image B has a stronger continuum flux
than image A, 
iii) the flux ratio of the continuum is not wavelength-dependent.

\subsection{PG~1115+080}

PG~1115+080 is a lensed QSO with a redshift of 1.72. The time-delay between images B and C of $23\pm 3.4$ days was initially 
measured
by Schechter et al. (1997) whereas a later analysis  by Barkana (1997)
resulted in a value of $25\pm 2$ days.
For the purpose of applying our method we ignore the time-delay between
images A1 and A2.  
Impey et al. (1998) had reported an anomalous flux ratio between images
A2 and A1.
Their observations ruled-out differential
extinction and microlensing as explanations for the anomalous flux ratio,
because of a lack of variability. 
They conclude that since an expected flux ratio between images A2 and A1 
of $\sim$ 1  is a generic feature of  the large-scale potential near a
fold caustic (from a simple lens model), only a
potential perturbation intermediate between that produced by { 
stars in the lens galaxy} (microlensing) and by the overall galaxy can
explain the anomalous flux ratio, and that the potential
of PG~1115+080 must be
perturbed either by a satellite galaxy or a globular cluster, i.e. that
millilensing is present.

Our search of the HST database found an observation of PG~1115+080 made on
1995 June 07, which covers the wavelength range from 850 \AA\ to 1200 \AA\
(rest wavelength) and observations made on 1996 January 21 (image A1) and
on 1996 January 24 (image A2) that cover the wavelength range  from 850 to  1750 \AA\ (rest
wavelength). These observations were performed with the  G270H and G400H gratings. More details describing these observations can be found in Bechtold et al. (2002). 

First, we calculated the indicator of variability ($\sigma$) and intristic
variability ($\sigma_I$) between observations made on 1996 January 21
(image A1) and on 1996 January
24 (image A2) for the wavelength range 
850 \AA\ -- 1200 \AA.  As shown in Figure 6a, the variability indicator 
is not significant, but it appears that a low level of variability 
may be present ($\sigma\approx 0.1$). 
In Figure 6b we show that weak intrinsic variability 
may be present ($\sigma_I\approx 0.05$).
 However, both indicators of variability are negligible with respect to
the level of
the rms noise ($\sim 0.3$). 

To further investigate the significance of variability, we 
determined the flux ratios between the same images at two epochs.
As shown in Figure 7, variability is not significant. Specifically the continuum 
flux of image A1 increased by about 3\%, and that of image A2 decreased by about 7\%  
between July 1995 and January 1996. (see Figure 7), however, both variations are
insignificant relative to the level of the rms noise.

We have also calculated the flux ratios between images A1 and A2
for the July 1995 and January 1996 observations (see Figures 8). 
As shown in Figure 8 the flux ratios tend to be slightly different
for different epochs, not only in magnitude, but also in the shape of the wavelength dependence. This result should be
taken with caution because of the high level of noise, especially in the spectra
obtained on 1996 January. We also find that the emission lines in the
spectrum of image A2 were stronger than those in image A1 during the July 1995 and January 1996 observations.
This difference in line strength is more obvious in the flux 
ratio between images A1 and A2 of
PG~1115+080 for the observation made on 1996 January 
when plotted in the spectral range of 850 \AA\ -- 1750 \AA 
(see Figure 9). The N III$\lambda$990.98, O IV$\lambda$1033.82,
Ly$\alpha$, Si IV$\lambda$1396.76 and C IV$\lambda$1549 lines 
are stronger in image A2 than in image A1. We note that
the flux ratio as defined in Eqs. (6) -- (7)
includes both line and continuum components.
Consequently, the difference of the intensity of the emission lines 
in images  A1 and A2 of PG~1115+080 as shown in Figures 8 and 9 may arise
in part from the different magnifications of the underlying continua in 
images A1 and A2.

We assumed that the relative macro-magnification of images A1 and
 A2 is achromatic (i.e.,  case II, Table 1). To search for possible
differences
between the magnification of the spectral lines and continuum, we
scaled the spectra observed in January 1996 of both images on the same
continuum level.
In Figure 10a we compare the flux vs. wavelength of images A1 and
A2  of PG~1115+080  after scaling the continuum of image A2 to that of A1
 (multiplying the flux of  
the image A2  continuum  by the function
$R(A1,A2)=2.75-5.9\cdot10^{-4}\cdot\lambda$).
We notice that the spectrum of image A2 has stronger lines (with respect to
the local continuum) than those of image A1. This is expected
if we have microlensing or
millilensing of the continuum. But this difference
cannot be explained by microlensing since the indicator of
variability is small with respect to rms noise ($\sigma\approx 0.1$). The
difference between the 
continuum  and line amplifications 
is more likely the result of millilensing. Here we should mention  that
long-term microlensing may produce a similar effect (case 
IVb), even with the indicator of variability being negligible in this
case, 
we cannot completely exclude long-term microlensing.

\begin{figure}
\includegraphics[width=8.5cm]{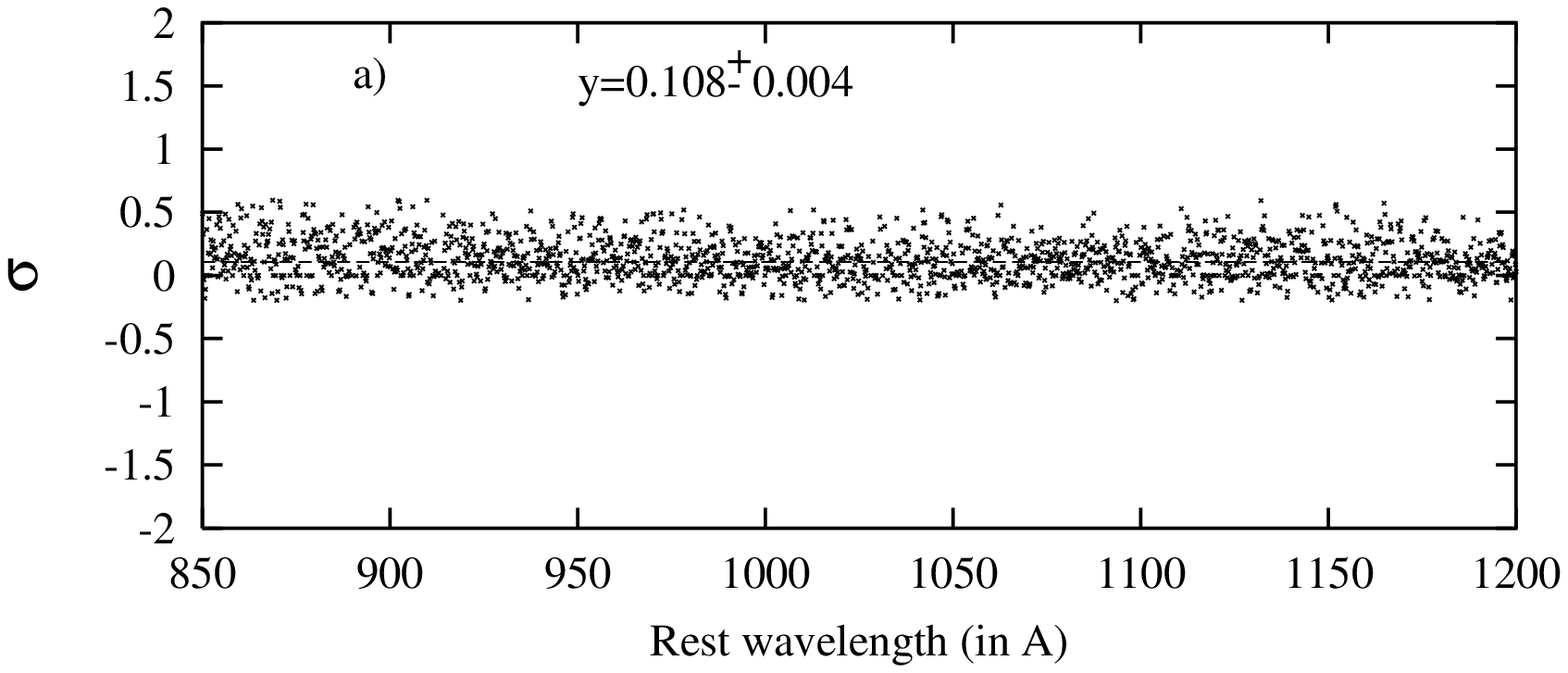}
\includegraphics[width=8.5cm]{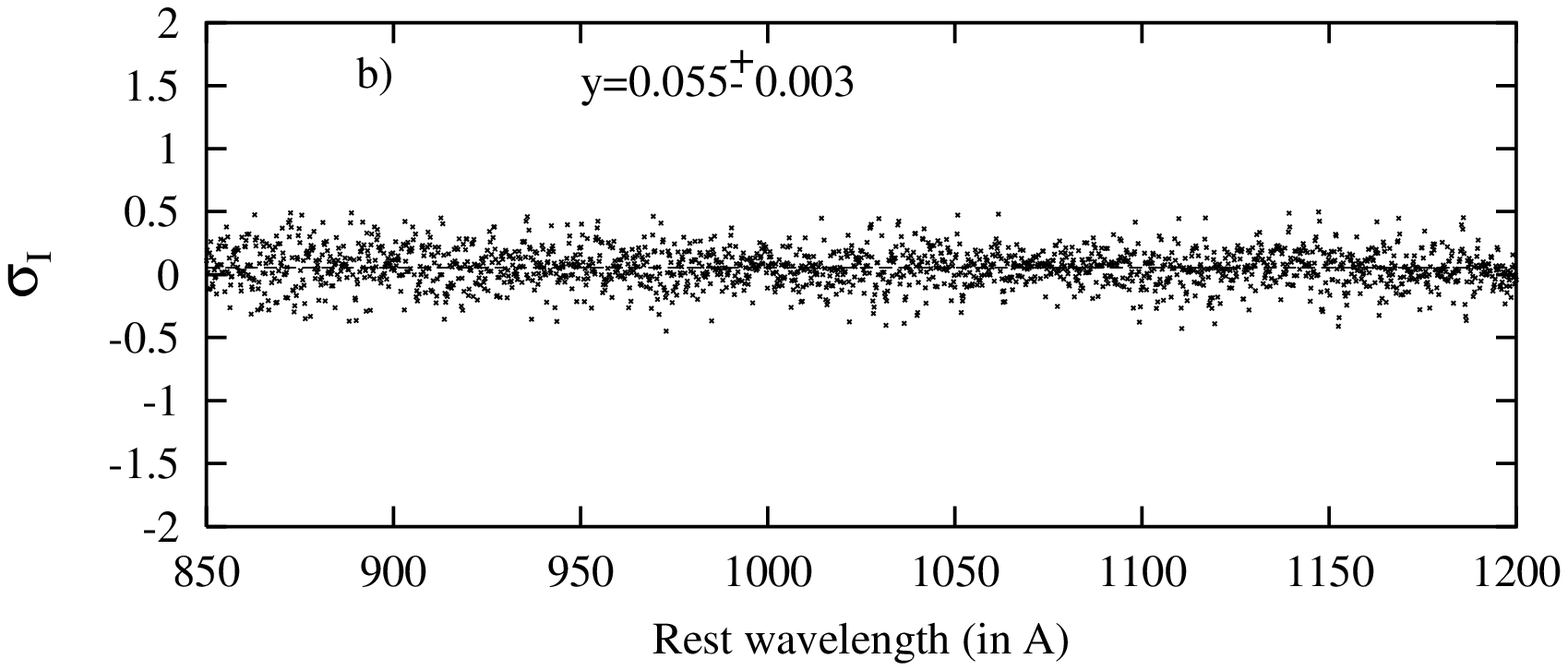}
\caption{a) The variability indicator ($\sigma$) and b) intristic
variability
indicator ($\sigma_I$) between images A1 and A2 of PG 1115+080}
\end{figure}

\begin{figure}
\includegraphics[width=8.5cm]{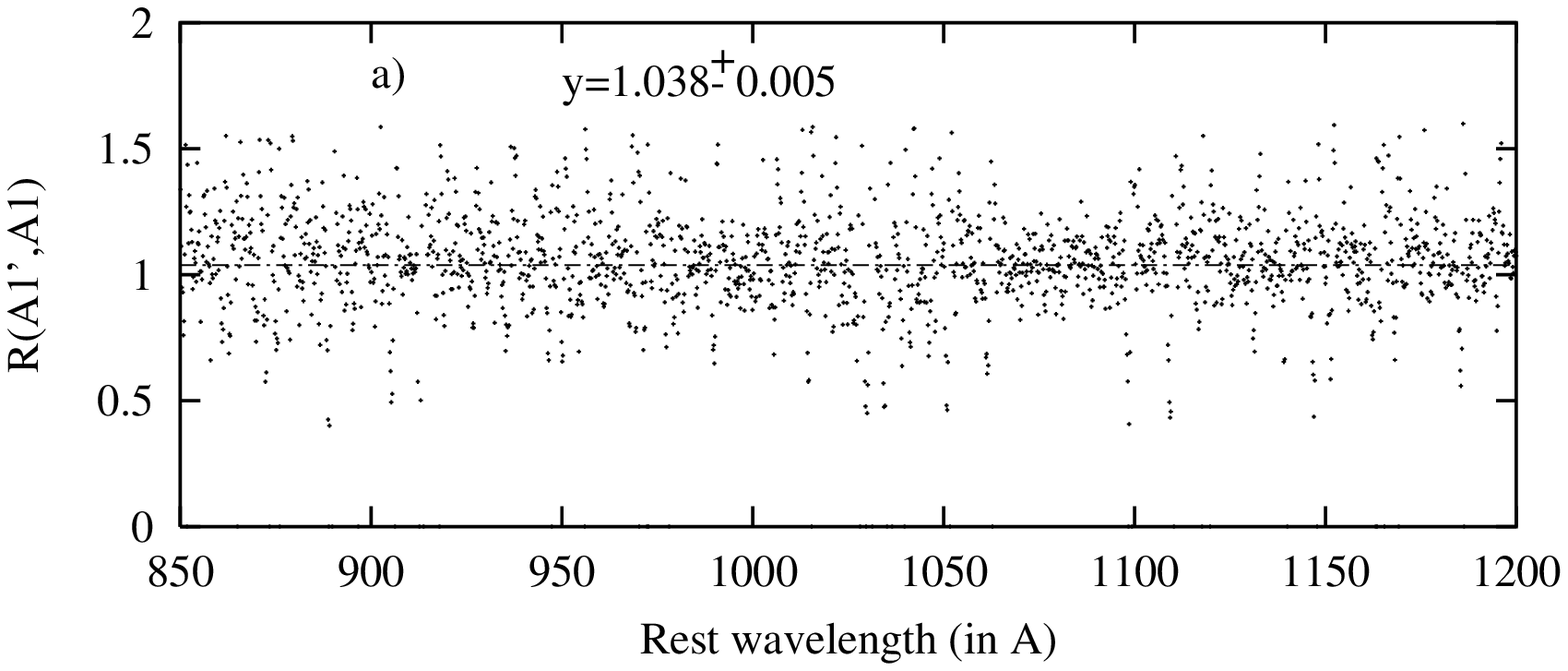}
\includegraphics[width=8.5cm]{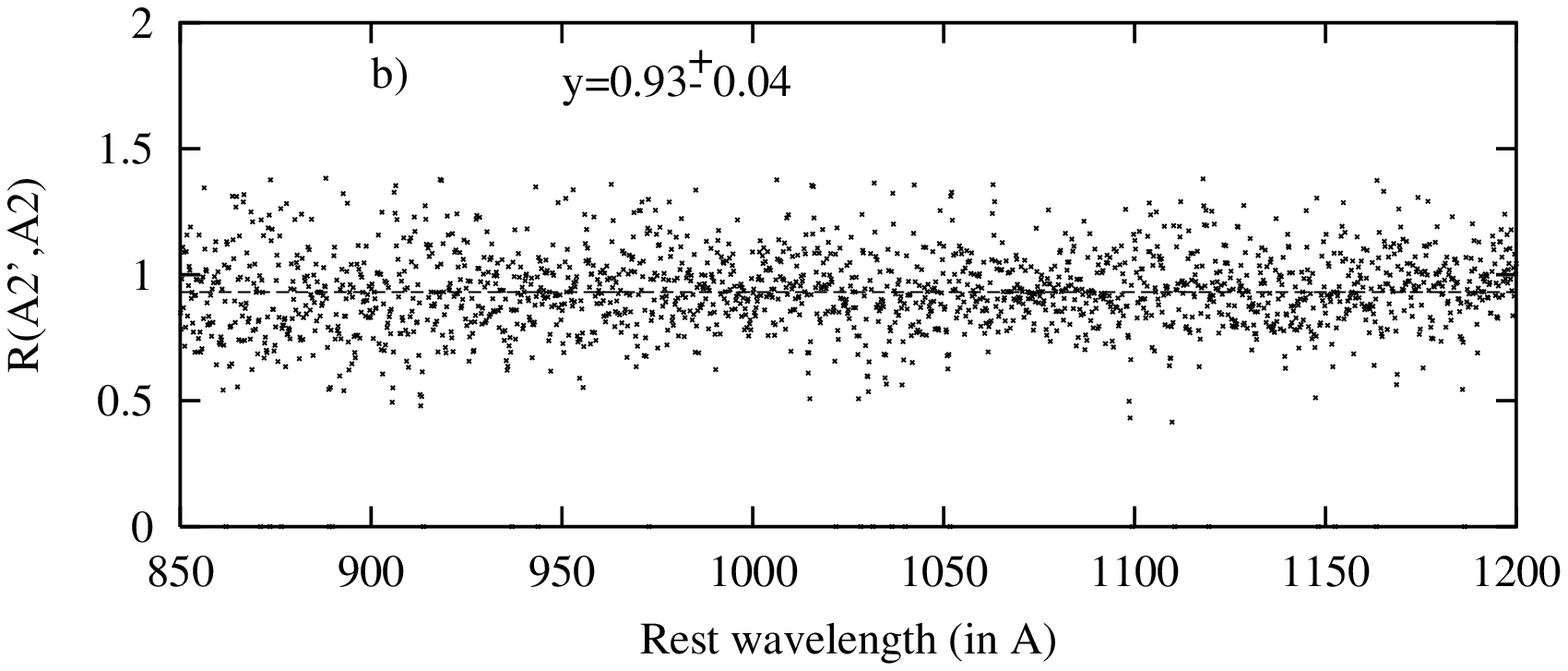}
\caption{The flux ratio between epochs 1995 July and 1996 January for
a) image A1, and b) image A2.}
\end{figure}

\begin{figure}
\includegraphics[width=8.5cm]{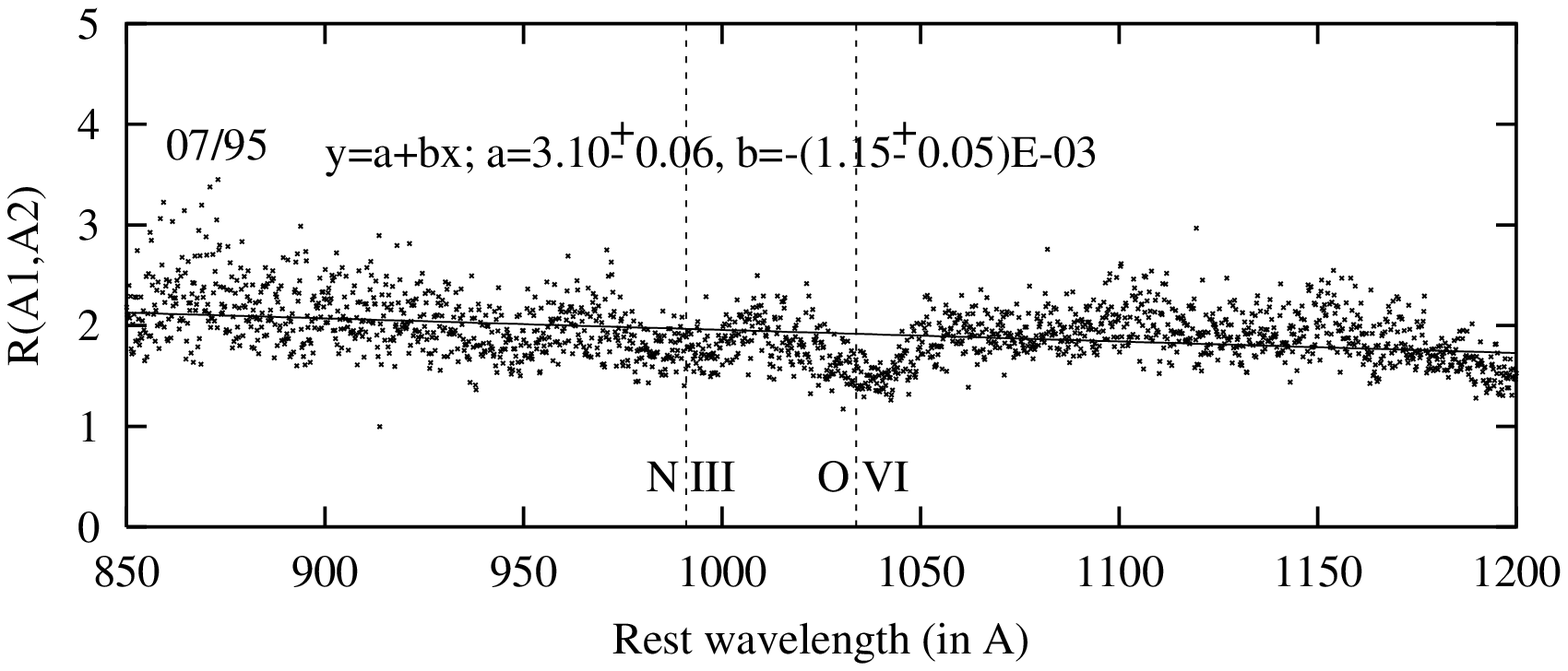}
\includegraphics[width=8.5cm]{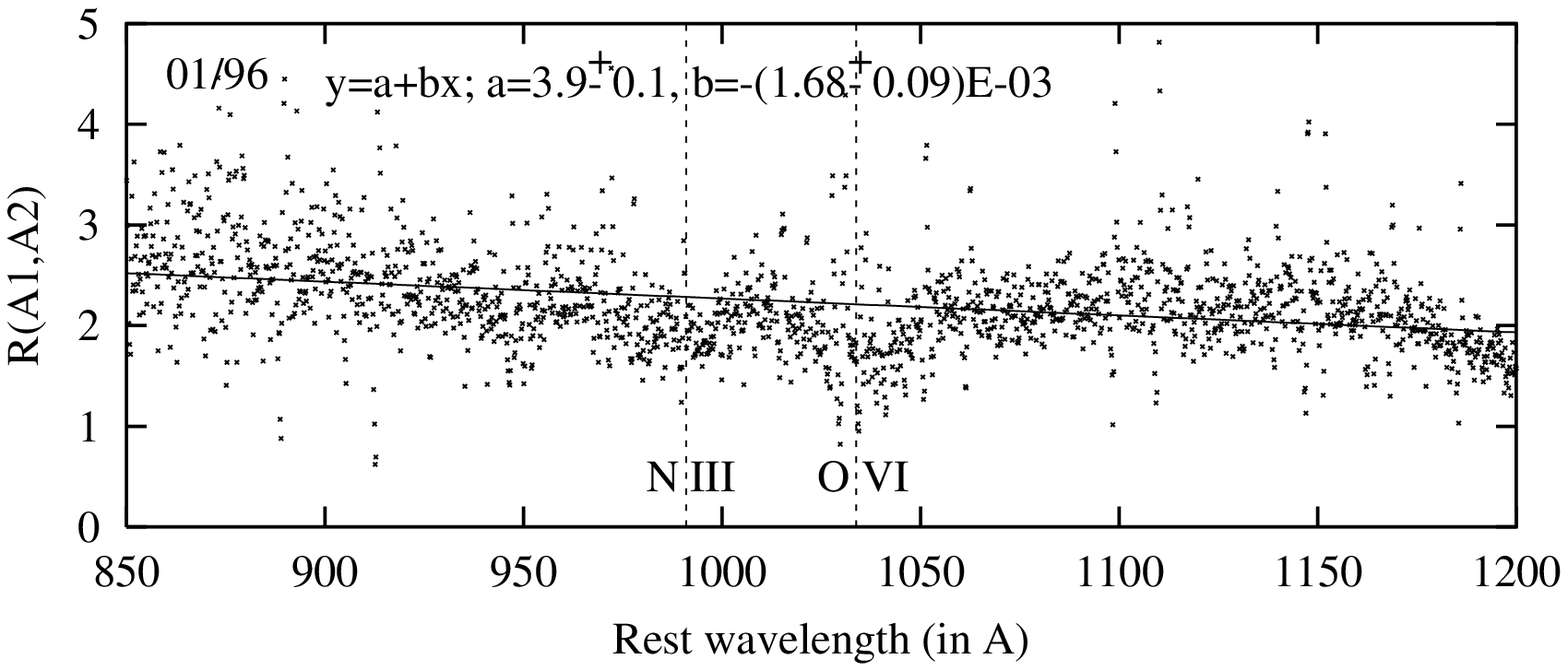}
\caption{The flux ratio as a function of wavelength between images
A1 and A2 of PG~1115+080 for two different epochs}
\end{figure}

\begin{figure}
\includegraphics[width=8.5cm]{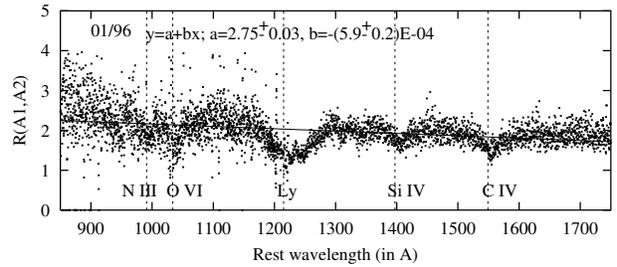}
\caption{The flux ratio of images A1 and A2 of PG~1115+080
 in the wavelength band of 850 -- 1750 \AA}
\end{figure}

\begin{figure}
\includegraphics[width=8.5cm]{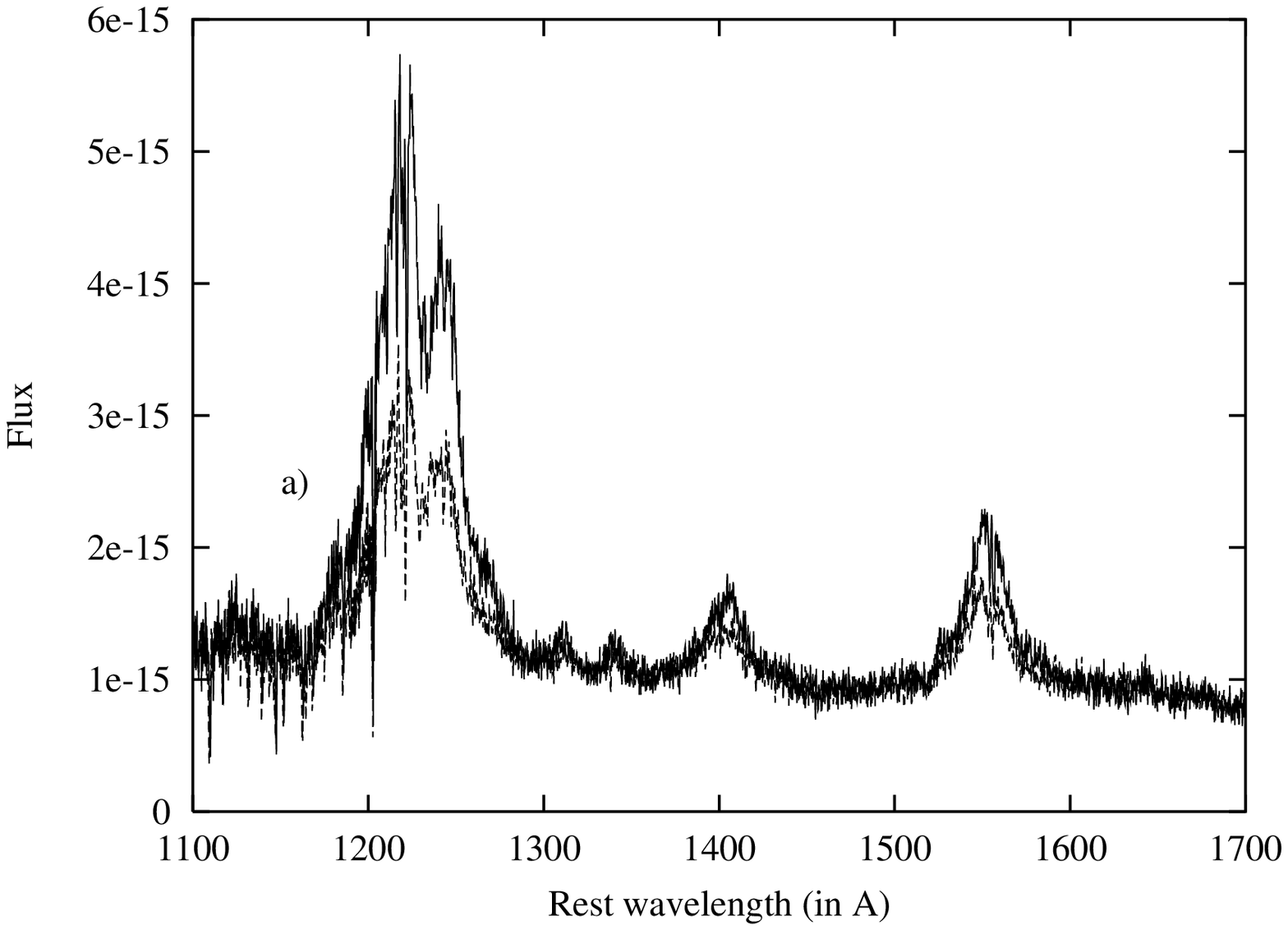}
\includegraphics[width=8.5cm]{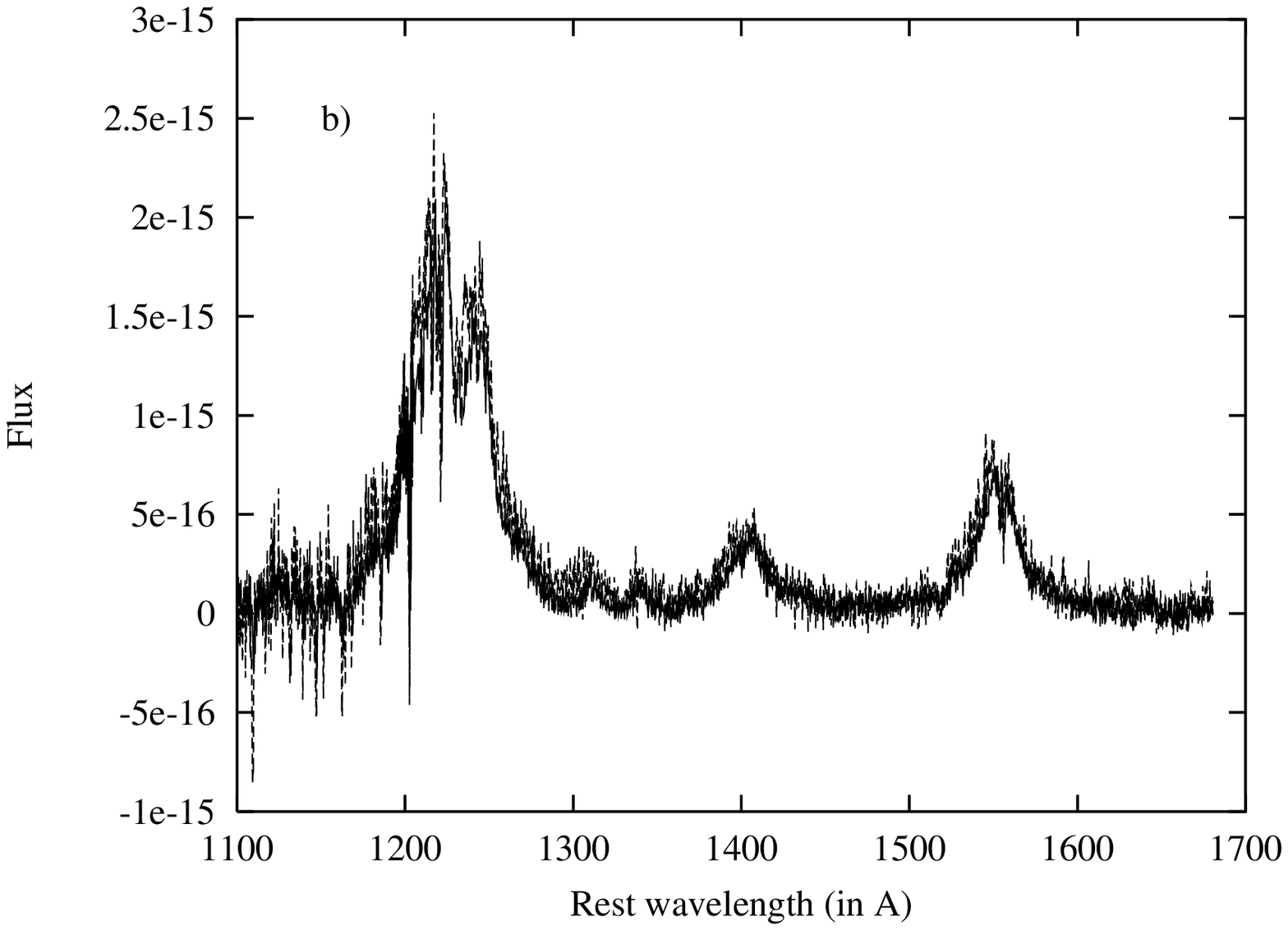}
\caption{The flux (given in $\rm erg\ cm^{-2}sec^{-1}A^{-1}$ units) of
image A1 of PG~1115+080 (dashed line) compared to 
that of image A2 of  PG~1115+080 (solid line): a) after scaling the
spectra of both images to
the same continuum (the continuum level of image A1),
and b) after subtraction of the continuum in both images. 
}\end{figure}

\begin{figure}
\includegraphics[width=8.5cm]{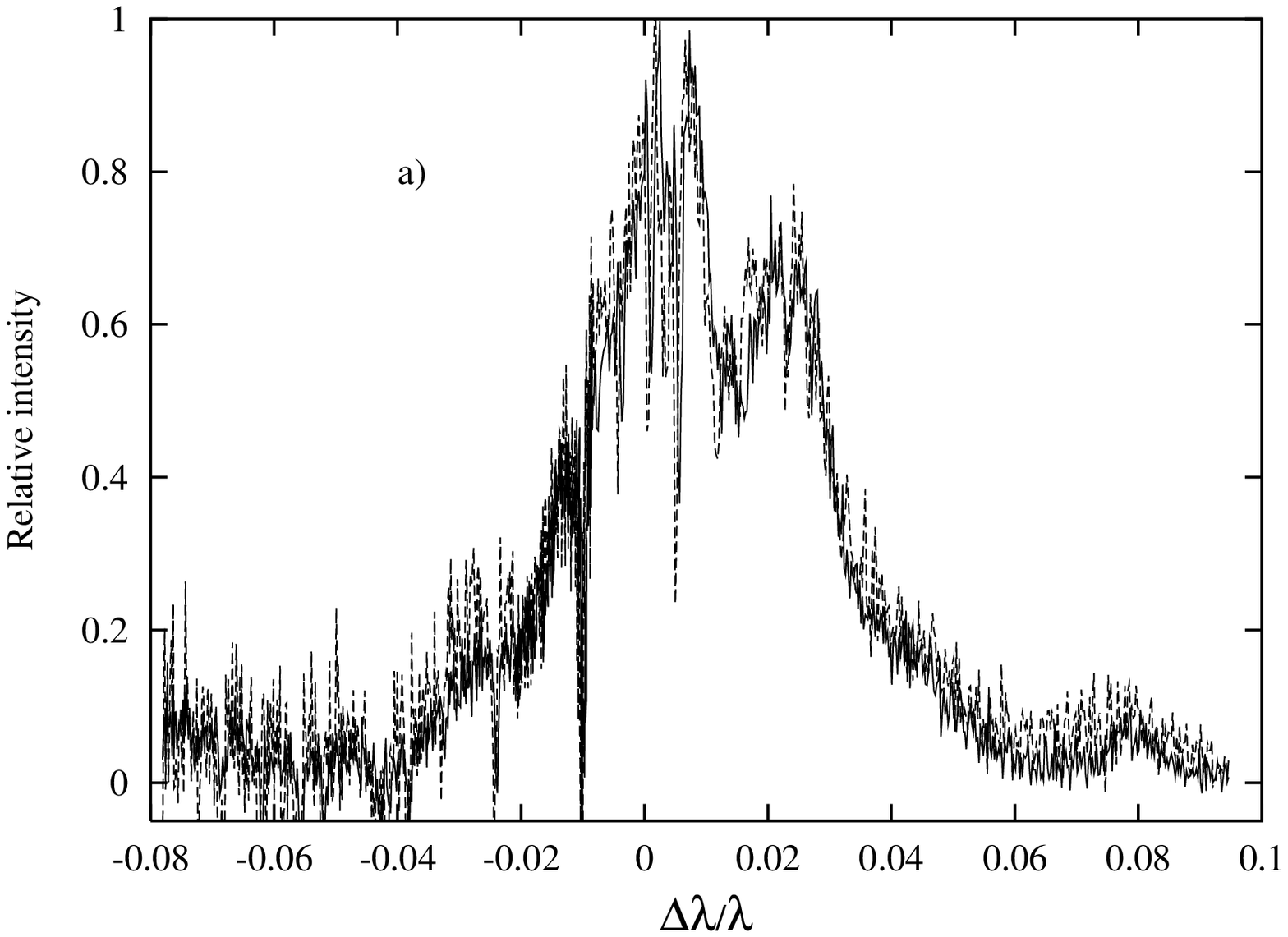}
\includegraphics[width=8.5cm]{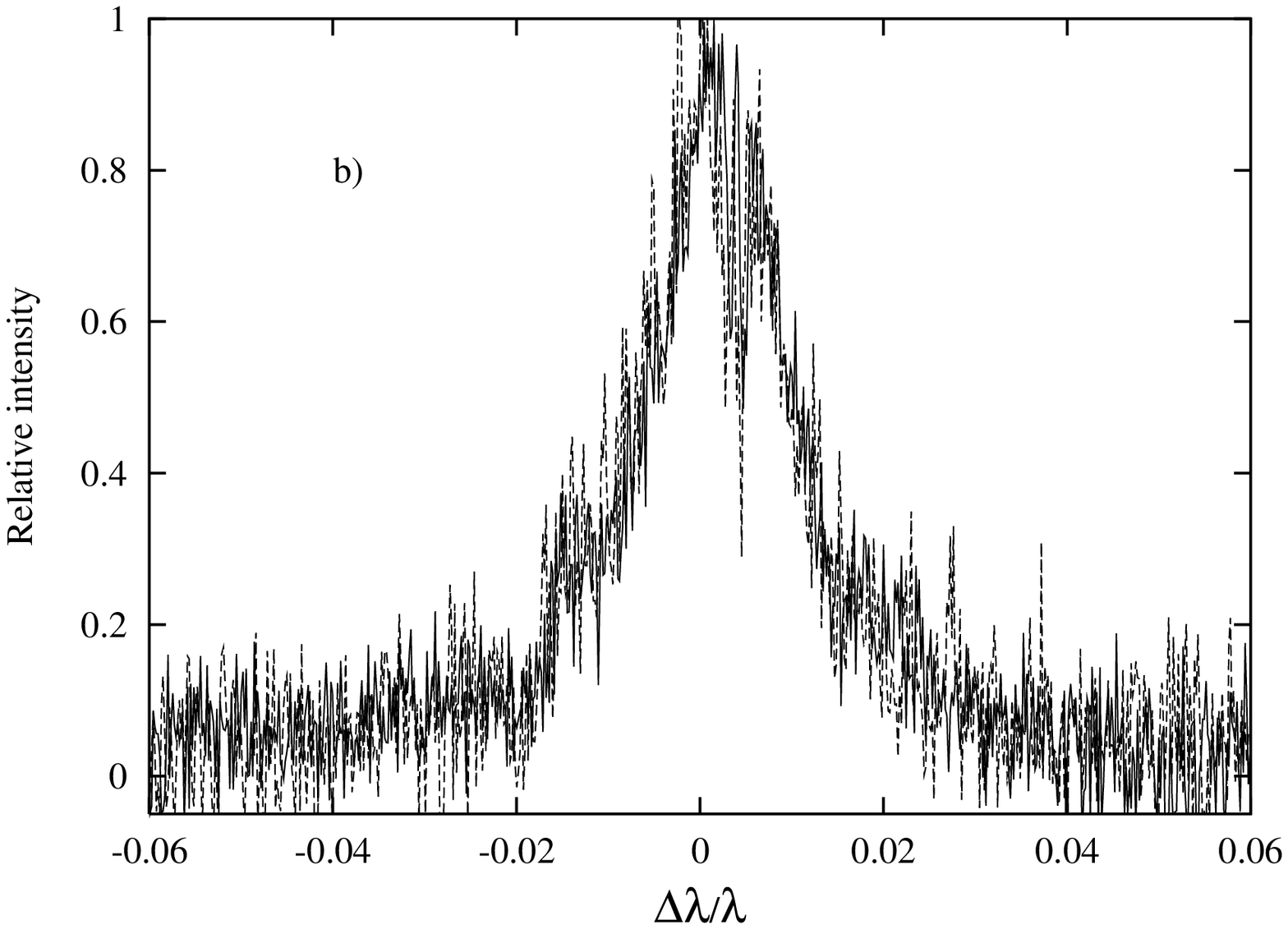}
\includegraphics[width=8.5cm]{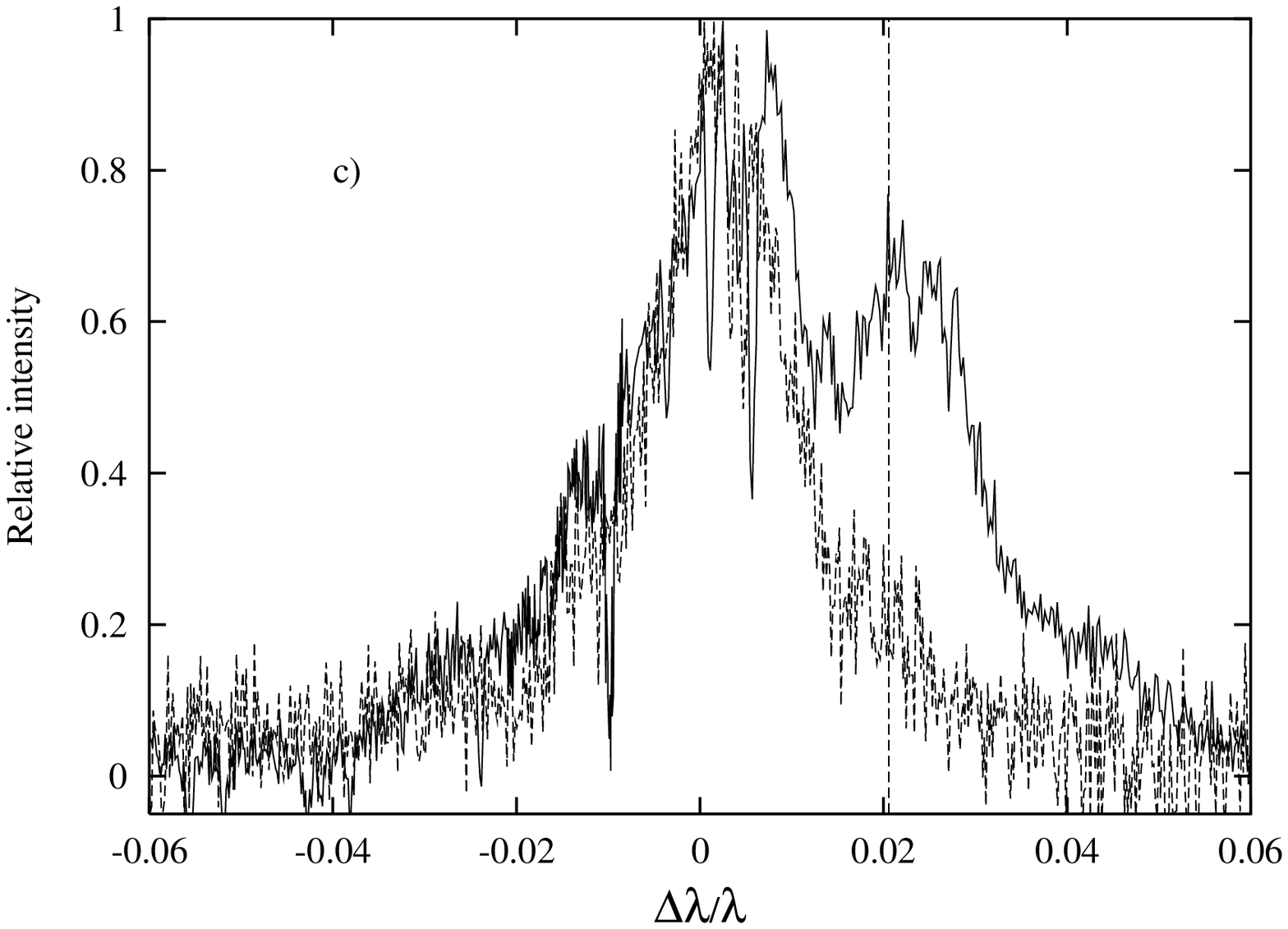}
\caption{A comparison between  : a) the Ly$\alpha$ line profiles of images
A1 and A2 of PG~1115+080;
b) the C IV line profiles  of images A1 and A2 of PG~1115+080; 
c) the Ly$\alpha$ and C IV line profiles of image A2 of PG~1115+080
(the vertical dashed line shows the position of the N v$\lambda$1240 line).
}\end{figure}

If we assume that the indicator of variability is negligible ($\sigma\sim
0$) and millilensing is present (case (2) of \S 2.1), then we expect
millilensing to cause different magnifications of the continuum and the
emission line
components, as indicated in Eq. (3).  
Since the line and continuum emission
components are additive (Eqs. 1-3), we subtracted the continuum
in both images (Eq. 11) and compared the line fluxes. 

A comparison of the line fluxes in the spectra of images A1 and A2
with subtracted continua is shown in Figure 10b.
A small differences in these fluxes can be
due to uncertainties in the subtracted continua, but we can conclude that 
the A1/A2 flux ratio of the lines  is  $(A1/A2)_L\sim
1$. This ratio corresponds to the expected value of $\sim 1$ if no micro
or millilensing were present 
because the 
 images are symmetrically arranged near a fold caustic (see Schneider et
al. 1992). On the other hand, the ratio of A1/A2 continuum flux is
wavelength dependent and in the observed wavelength range (from 850 \AA\
to
1750 \AA ) can be approximated by the function $R(A1,A2)_{\rm
continuum}\approx
2.75-5.9\cdot 10^{-4}\cdot\lambda$, i.e. the amplification in  the continuum is
changing from $(A1/A2)_{\rm continuum}\approx 2.3$ at 850 \AA\ to
$(A1/A2)_{\rm continuum}\approx 1.75$ at 1750 \AA .

Since millilensing is similar to
 microlensing, we may expect the line profiles of images A1 and A2 to be
different
(Popovi\'c et al. 2001, Abajas et al. 2002).
In Figure 11 we compare the Ly$\alpha$ and C IV$\lambda$ 1549 line
profiles between
images A1  and A2. We also show a comparison between the Ly$\alpha$ and C
IV$\lambda$ 1549 line profiles
from the spectrum of image A2 since the
line emitting regions of different emission lines may be different
(e.g., Wandel et al. 2000; Kaspi et al. 2000).

For the purpose of these comparisons, we normalized the fluxes of the 
Ly$\alpha$ and CIV$\lambda$1549 lines
to one and converted wavelengths to velocities :  $\lambda \to
X=(\lambda-\lambda_0)/\lambda_0$.
As shown in Figure 11 there are no significant differences between the line profiles of the
same lines in the spectra 
of images A1  and A2. These comparisons suggest that millilensing 
does not affect the emission line region of PG~1115+080. 
 
In Figure 11c,  we note that the red wing of the Ly$\alpha$
line is stronger than the C IV red wing. We propose that this
difference is likely caused by the contribution of the NV$\lambda$1240 line. 
We note that such a strong 
NV$\lambda$1240 doublet ($F_{N\ V}\sim 0.5 F_{Ly}$) is  unusual for QSOs,
with a more typical value expected to lie in the range
$F_{N\ V}\sim  (0.1-0.2) F_{Ly}$ (e.g., Laor 1994).

Our investigation of the spectra of images A1 and A2 of PG~1115+080,
confirms the anomalous spectral distribution and relative magnifications
of these two images, first noted by Impey et al. (1998).
The main results of our analysis of the spectra of PG1115+080 are
as follows: 
i) We only detect very weak variability in images A1 and A2 of PG~1115+080 
between two observations;
ii) the differential magnification of the line and the continuum
components as well as wavelength-dependent amplification
may be caused by millilensing. 
This is consisted with an earlier proposal 
by Impey et al. (1998) that suggested the presence of  significant
perturbations to the main
lens galaxy of the PG~1115+080 lens system 
by a satellite galaxy or globular cluster. 
iii) Millilensing affects the continua of A1 and A2 in observed wavelength
range
($F(A1)_C/F(A2)_C\approx
1.75-2.3$, expected $\sim 1$),
but it does not appear to
significantly affect the line profiles of the images ($F(A1)_L/F(A2)_L\approx
1$).

\subsection{QSO~1413+117 - Cloverleaf}  

The spectra of the images of QSO~1413+117 were observed  in
the spectral ranges of 900--1350 \AA\
and 1300--1800 \AA\  and an analysis of these spectra was 
presented in Monier et al. (1998). They found that the emission spectral
lines show an intrinsic absorption component 
and that the emission lines in the spectrum of image D 
are significantly weaker than in the other three images.

\begin{figure*}
\includegraphics[width=8.5cm]{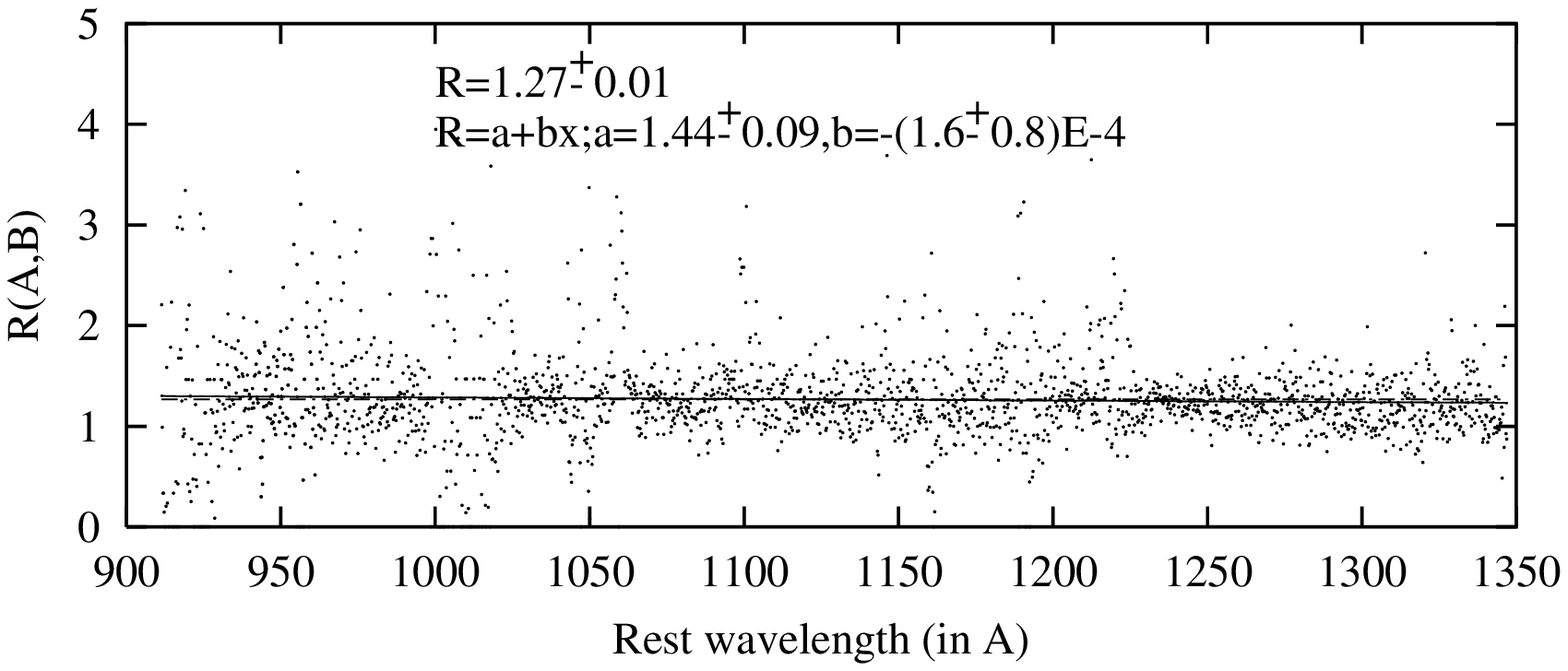} 
\includegraphics[width=8.5cm]{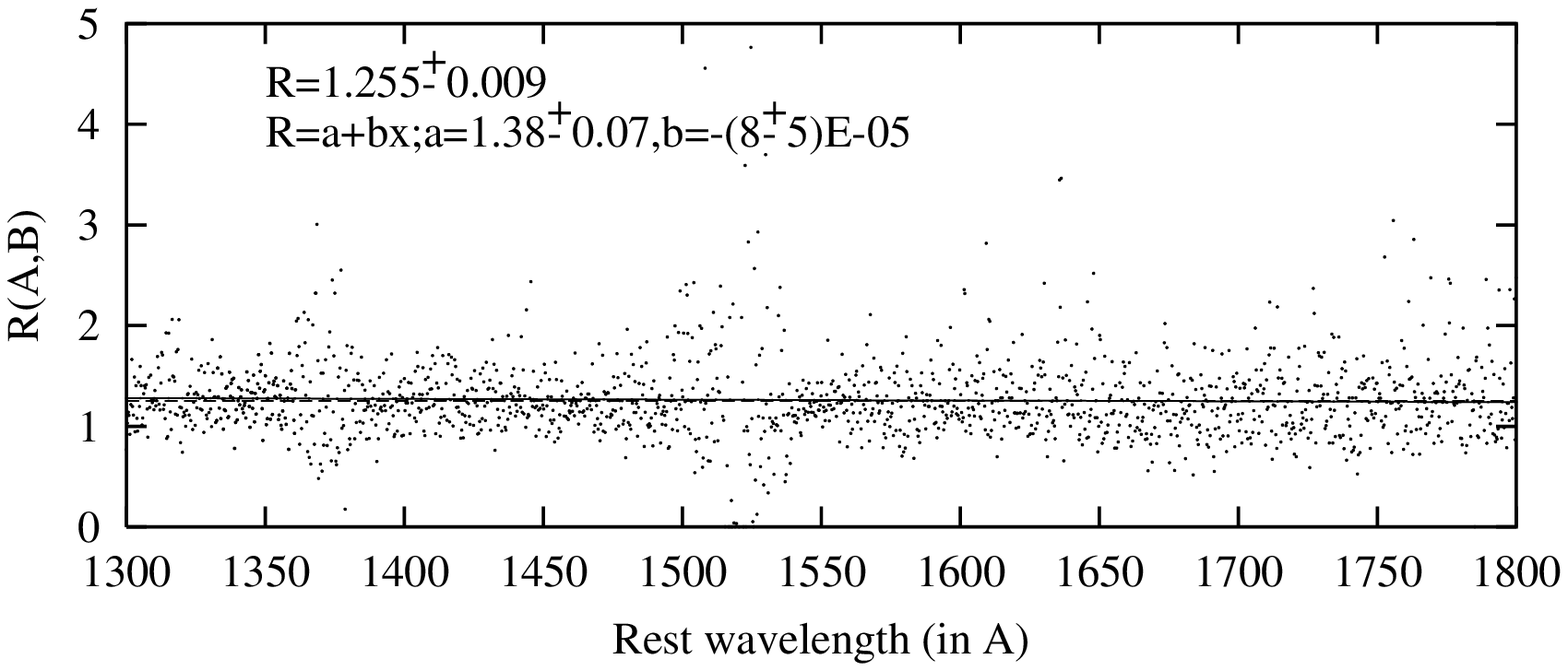}
\includegraphics[width=8.5cm]{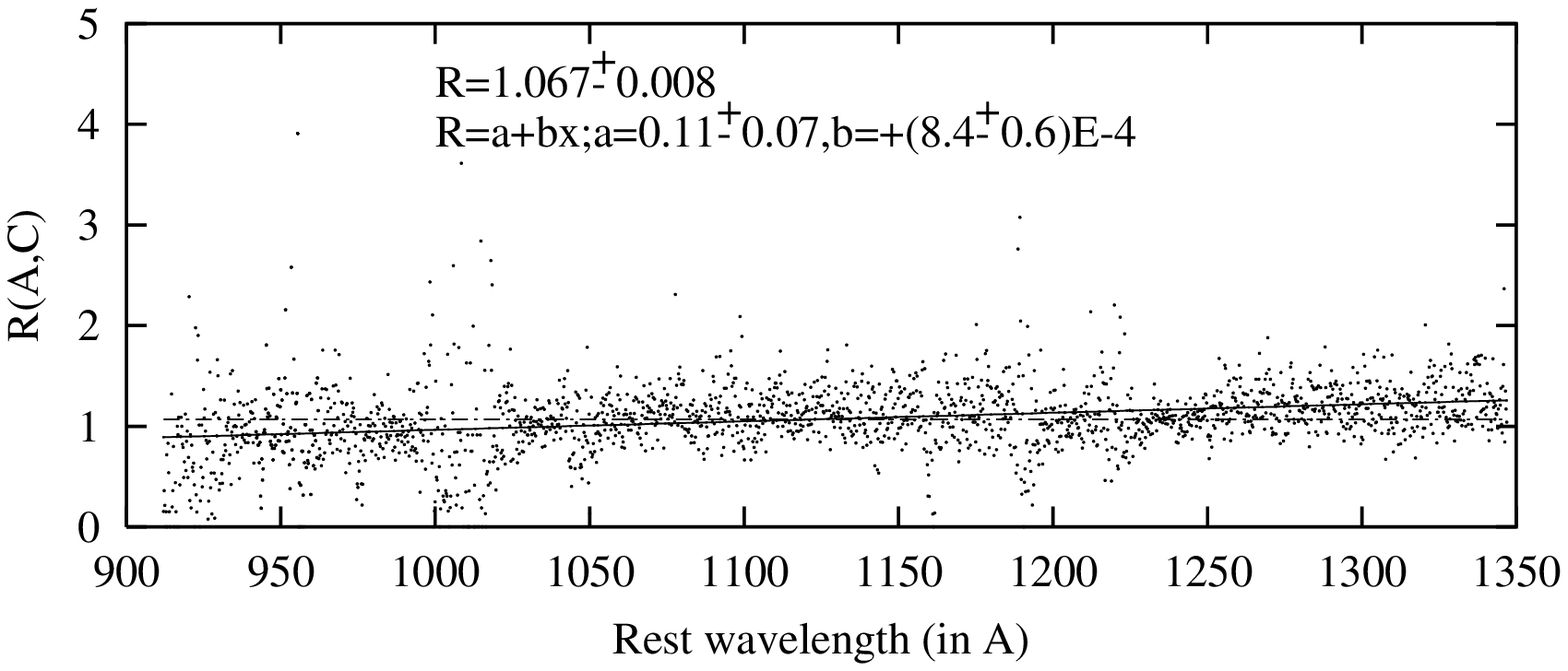}
\includegraphics[width=8.5cm]{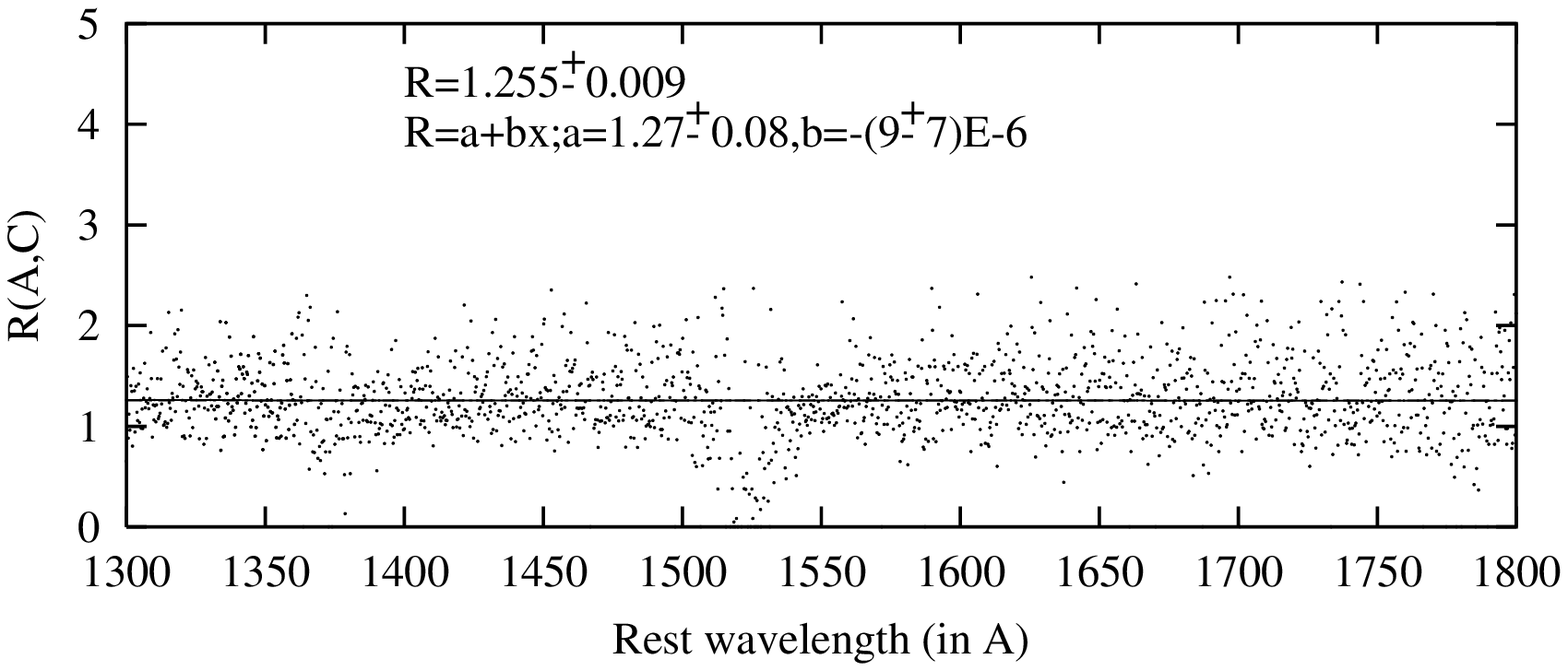}
\includegraphics[width=8.5cm]{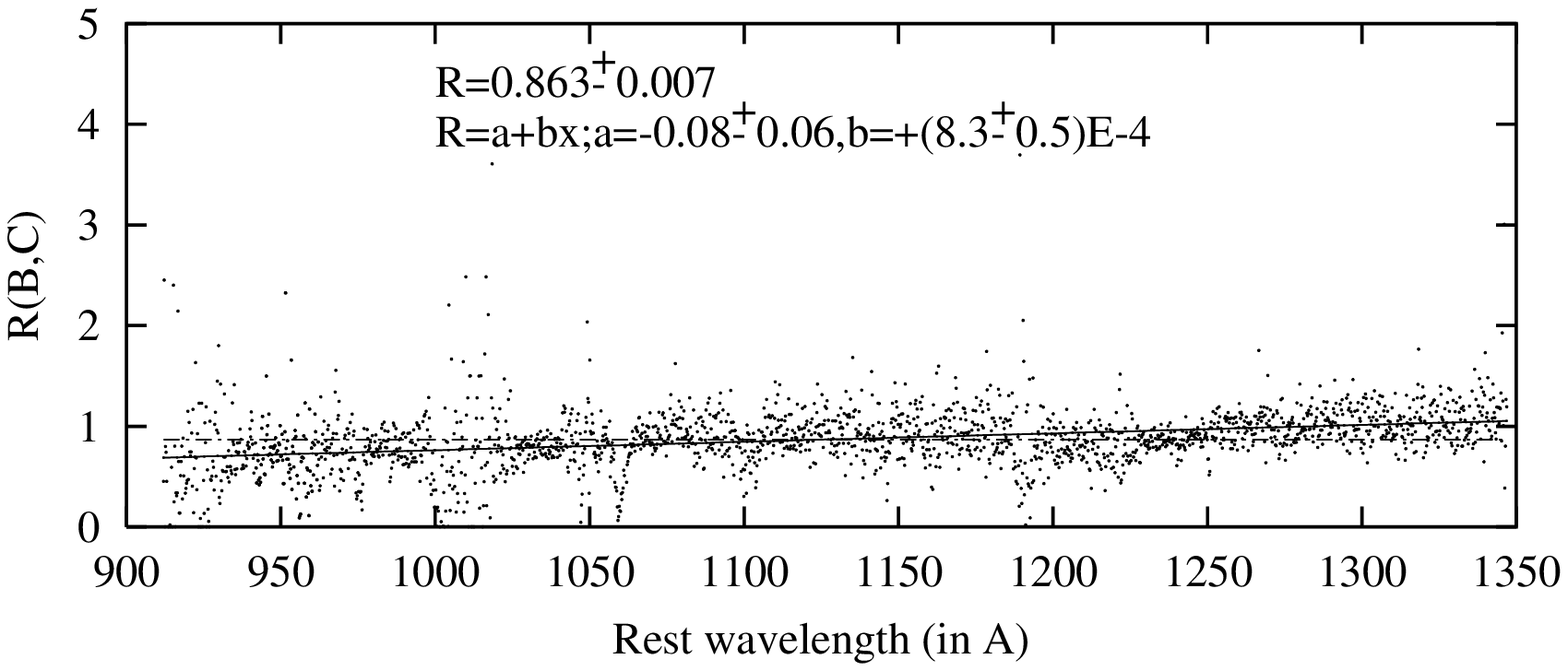}
\includegraphics[width=8.5cm]{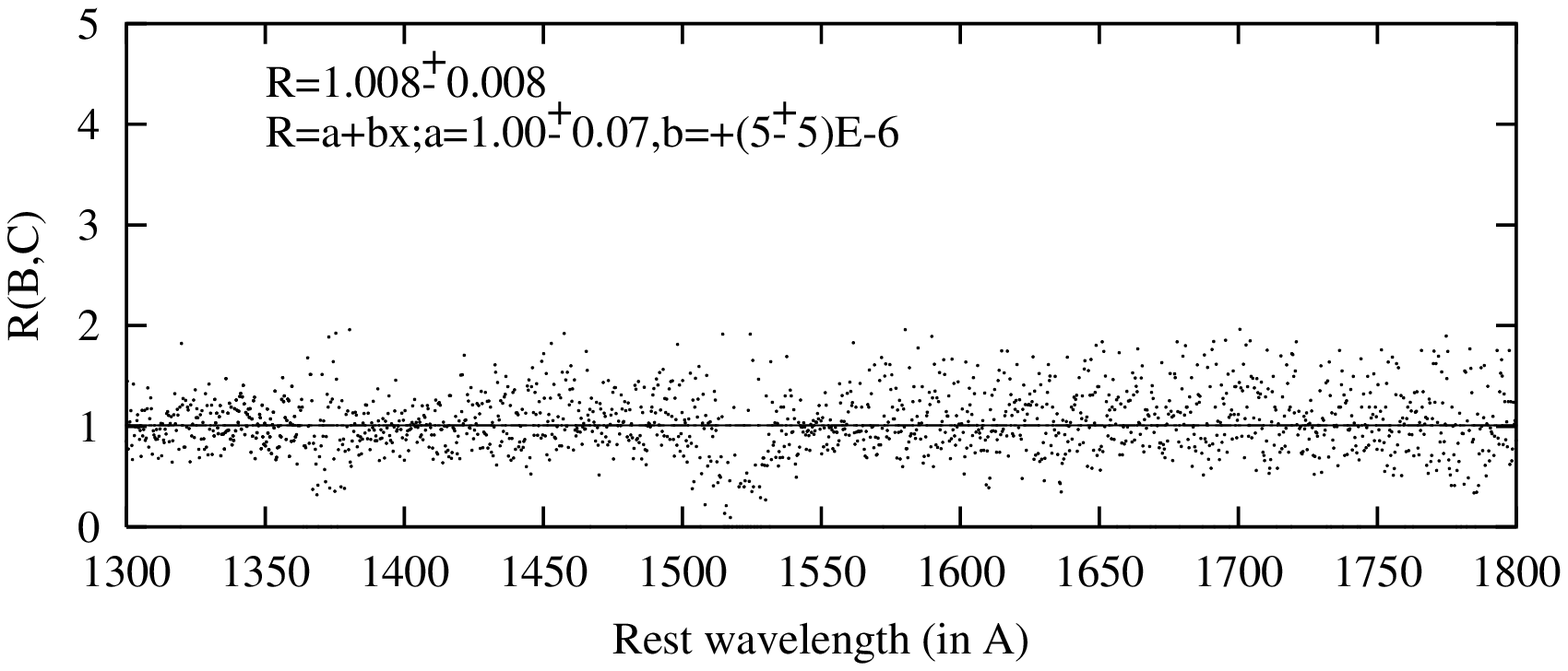}
\includegraphics[width=8.5cm]{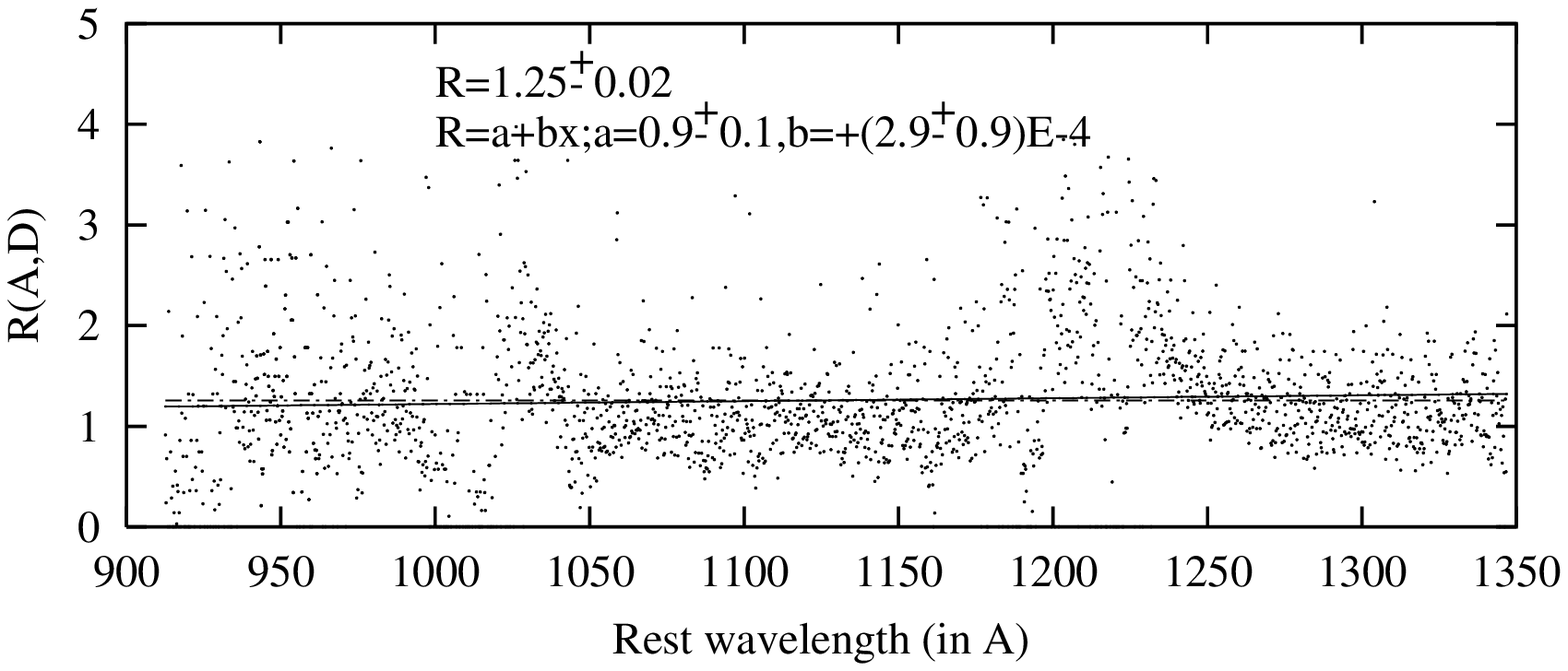}
\includegraphics[width=8.5cm]{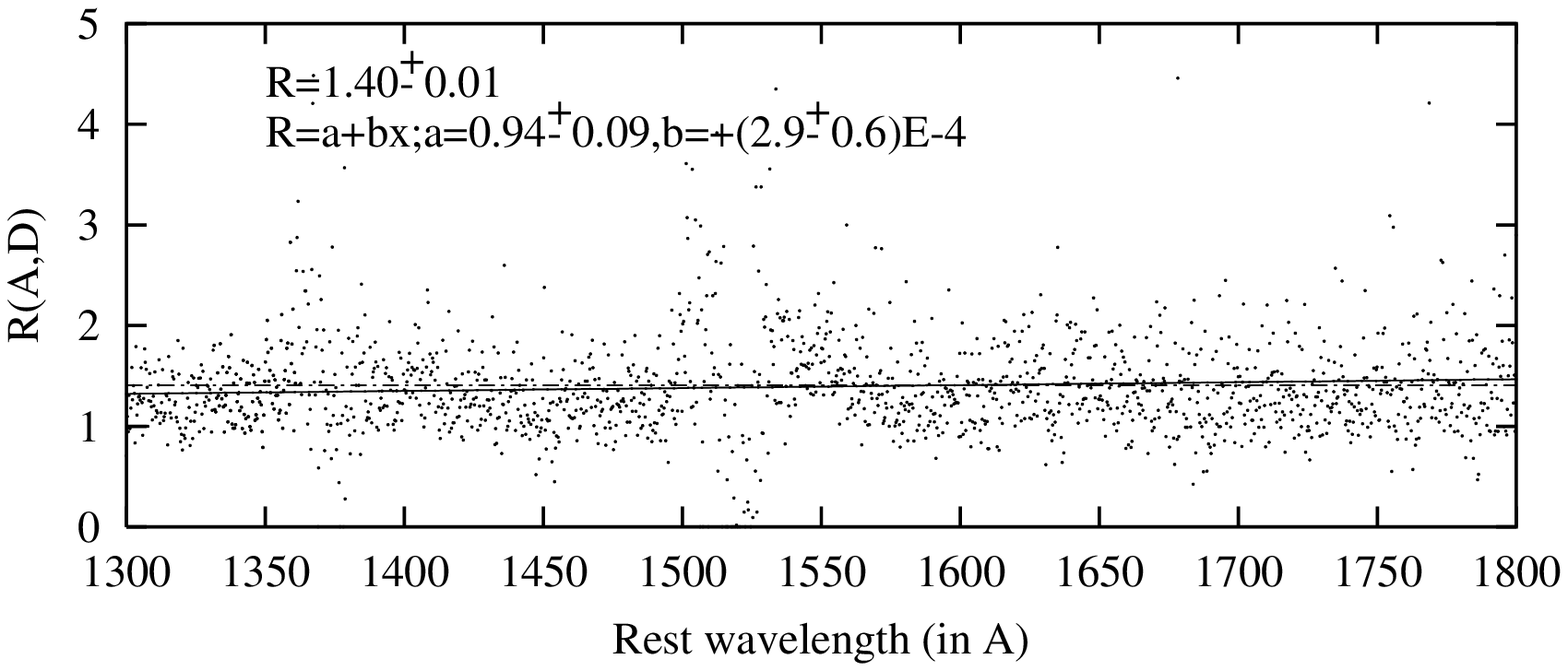}
\includegraphics[width=8.5cm]{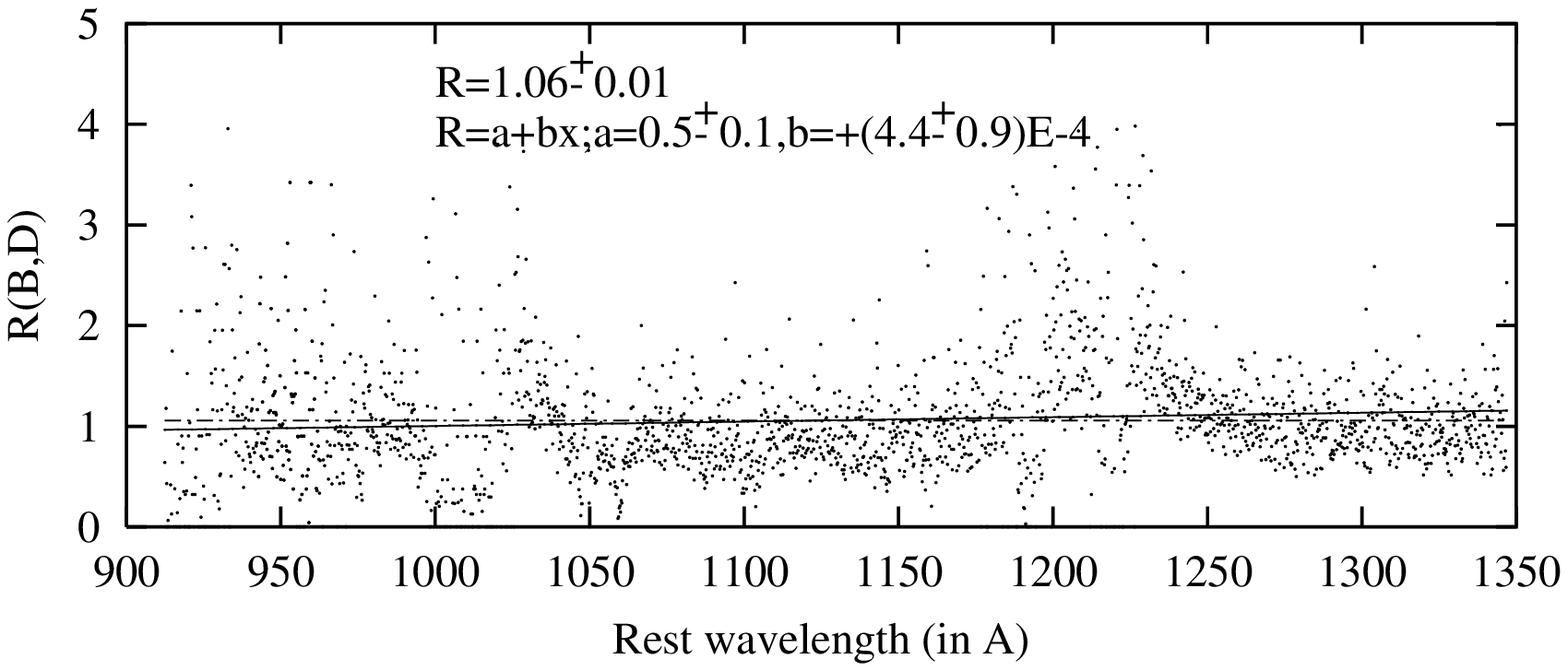}
\includegraphics[width=8.5cm]{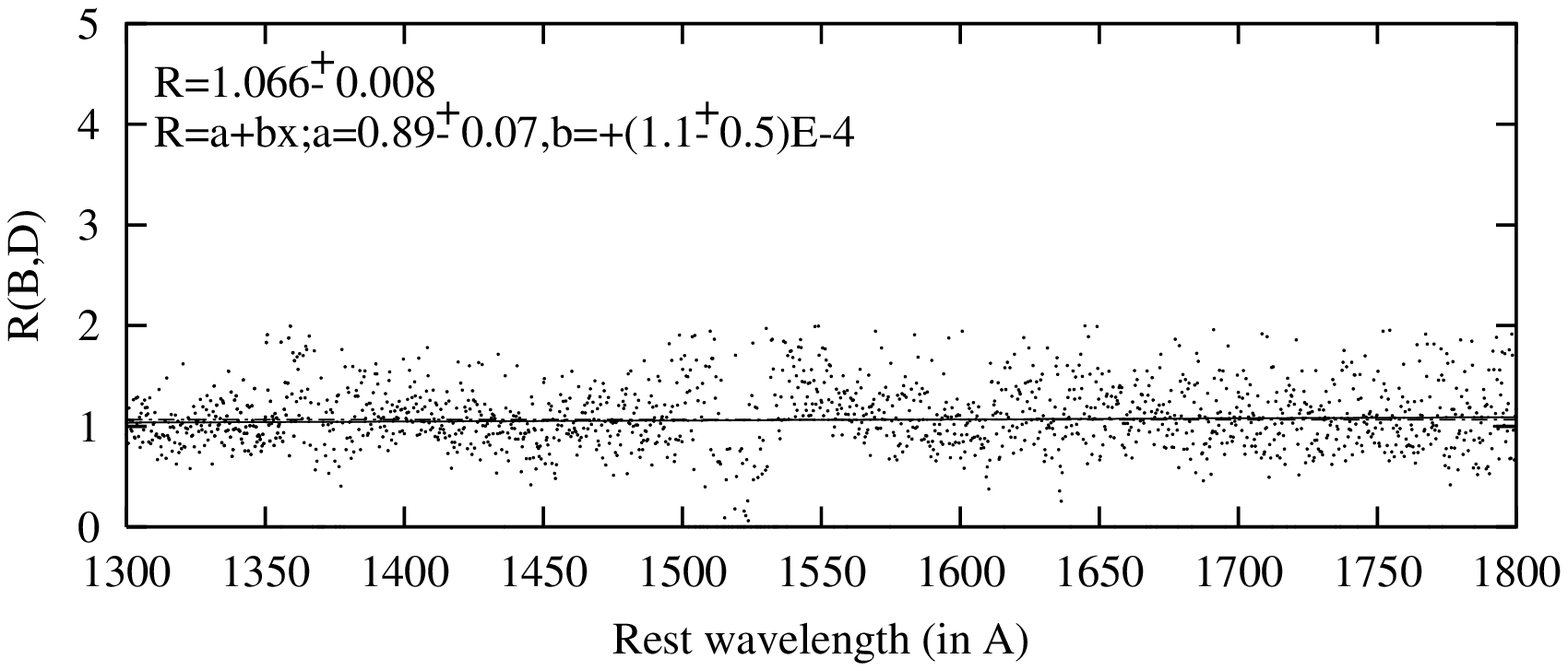}
\includegraphics[width=8.5cm]{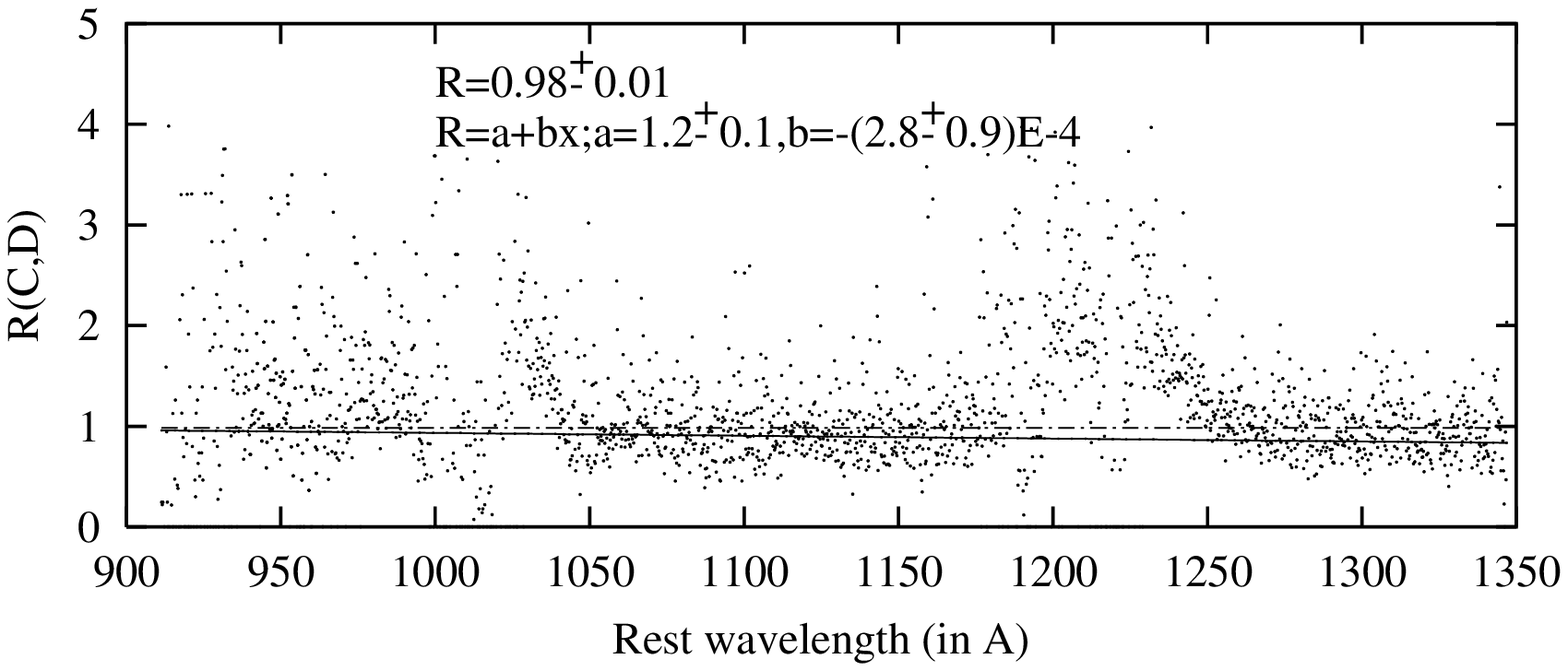}
\includegraphics[width=8.5cm]{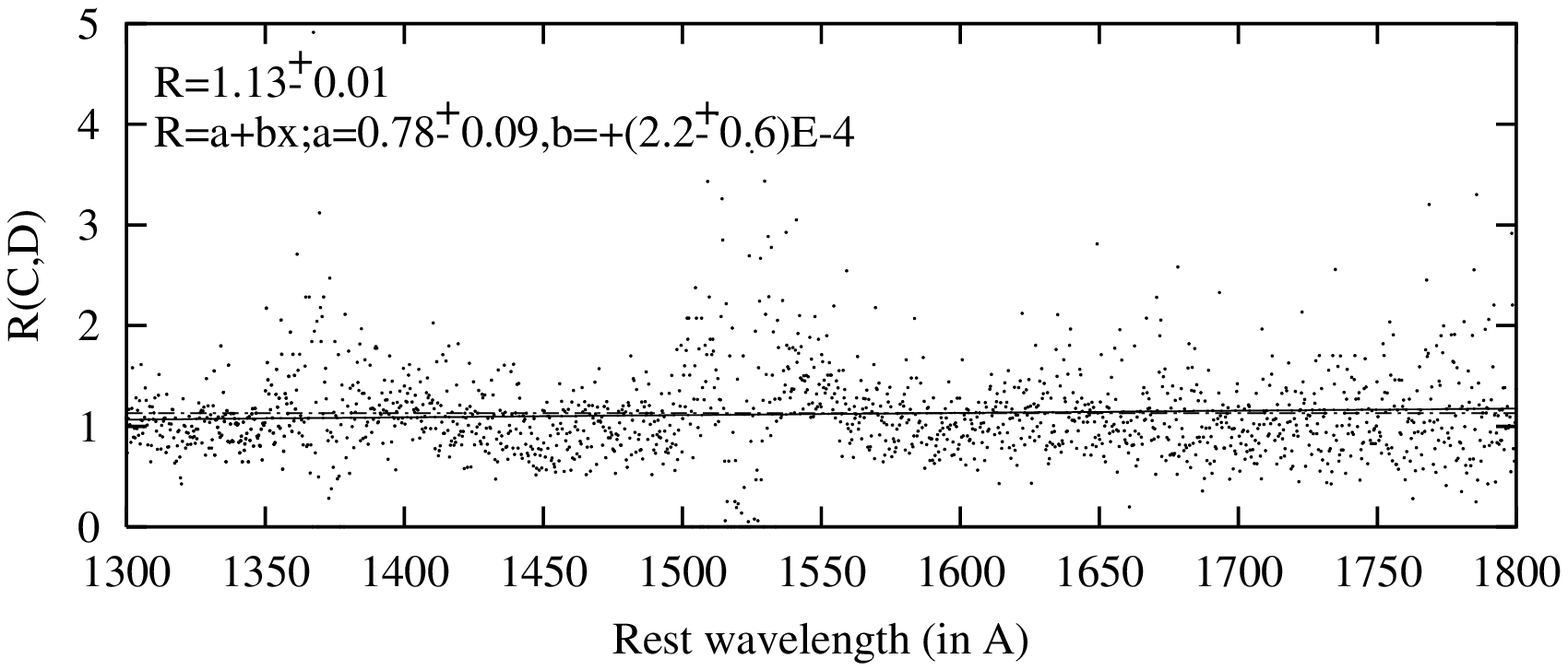}
\caption{ The flux ratio ($R_{i,j}$) of different images of QSO~1413+117
observed on the same date. 
Left panels: $R_{i,j}$ versus wavelength ranging from 900
\AA\ to 1350 \AA. Observations were made on 1994 December 24.
Right panels: $R_{i,j}$ versus wavelength ranging from
1300 \AA\ to 1800 \AA.
Observations of images A and D were performed on 1993
June 23 and of images B and C on 1993 June 27.
From top to bottom, we show the flux ratios between the following
images: 
A and B, A and C, B and C, A and D, B and D; C and D. The solid 
and dashed lines indicate the best fits of $R_{i,j}=a+b\cdot \lambda$ and  
$R_{i,j}=const$, respectively.} \end{figure*}

For our analysis we used the observations performed on 1993 June 23,
(images A and D) and 1993 June 27, (images B and C) covering the 
spectral range 1300 -- 1800 \AA\  and December 24,
1994 (all images) covering the spectral range 900 -- 1350\AA .
The number and separation in time of these observations are not 
ideal for an accurate application of our method, however, as we will show
we can still use our method to discuss the presence of a lensing event.   

In Figure 12 we show the flux ratios as a function of wavelength
(for both spectral ranges) of all images of QSO~1413+117.  We fit the
flux ratios with models of the form $R=a+b\cdot \lambda$ and $R=const.$

An inspection of the flux ratios indicates that:

i) during the first epoch (1994 December 24) only the ratios $R(A,C)$ and
$R(B,C)$ have a significant non-zero slope, i.e., they are 
wavelength-dependent. This might also be the case for $R(C,D)$, but due to
large residuals in the emission lines and the high level of noise of the ratio this result
should be taken with caution. In the other cases the ratios tend to be constant.

ii) the flux ratios $R(A,B)$, $R(A,D)$, and $R(B,D)$ show no 
significant differences between all epochs observed, but
flux ratios containing image C are different between epochs, i.e.,
$R(C,D)$, $R(B,C)$, and $R(A,C)$ show significant changes of the fitted slope 
between epochs.

iii) the spectra of image D for all observations show weaker UV
emission lines than those of the remaining images as was previous noted
in Monier et al. (1998).

From these observations we conclude that significant variability only of
image C is present, and considering that the ratios $R(A,C)$ and
$R(B,C)$ tend to be
wavelength-dependent, we suspect that 
microlensing of image C was likely present during the
observations.

\section{Conclusions}

In this paper we investigated the influence of gravitational lensing on
the spectra of lensed QSOs. Starting from the assumption that the
magnification due to a lens
is in general a complex function, (i.e., the presence of
globular clusters or satellite dwarf galaxies in the lens galaxy may
introduce
perturbations to the potential of the principal lens galaxy as noted by
Impey et al. 1998) and that the line and continuum emitting regions are
 different in
sizes and in geometries, we found that the
magnification of the
spectra of the different images may be chromatic
(as was noted in Wambsganss \& Paczy\'nski 1991; Lewis et al. 1998;
Wisotzki et al. 2003 and Wucknitz et al. 2003).

Here we briefly summarize several conclusions
that arise
from Eqs. (1-11) in \S 2:

i) in general the complex geometries of the emitting regions of QSOs   
and the complex potentials of the lens galaxies need to be taken into
account
when determining their influence on the observed spectra of the
continuum and line emission;

ii) microlensing events lead to wavelength-dependent magnifications
of the continuum that can be used as indicators of their presence;

iii) the average magnification due to microlensing
may be determined from the flux ratio of an image observed at
two different epochs separated by the time-delay and from the flux
ratio between different images
observed at the same time;

Here we propose a method to infer the presence of microlensing,
millilensing
and intrinsic variability in one or both of the images
of a lensed quasar. The method is outlined in the following steps:

(a) Obtain simultaneous observations of the spectra of the images of a
lensed quasar (at least) at two different epochs separated by the
time-delay.  

(b) Calculate the variability parameters $\sigma$ and $\sigma_I$ using
equations
(8) and (10), and determine the
flux ratios $R_{A'A}$, $R_{B'B}$, $R_{AB}$, $R_{A'B'}$, and
$R_{A'B}$.

(c) Use Table 1 to determine the possible presence of microlensing, 
millilensing or
intrinsic variability of the quasar.

Using our proposed method we investigated the influence of microlensing on the
FOS spectra of three lensed QSOs. 
We presented an example of the chromatic effect in our analysis of the FOS
spectra
of  PG 1115+080. In particular, the application of our lensing method to
this system
indicates that the line and continuum components of
the lensed QSO spectra of PG1115+080 have different magnifications
during the HST observations.

The application of our method to investigate the effect of lensing
on the spectra of lensed QSOs may be useful not only for 
constraining the unresolved structure of the central regions of QSOs,
but also for providing insight to the complex structure of the lens galaxy
and possible substructure in the lens galaxy.

\section*{Acknowledgments}

This work is a part of the project "Astrophysical Spectroscopy of
Extragalactic Objects'' supported by the
Ministry of Science and Technologies and Development
of Serbia and the Alexander von Humboldt foundation 
through the program for foreign scholars. We would like to thank the
anonymous referee for very useful comments.

{}

\end{document}